\documentclass{aa}

\usepackage{graphicx}
\usepackage{float}
\usepackage{longtable}
\usepackage{xcolor}
\usepackage{refcount}
\usepackage{txfonts}
\usepackage{lipsum}
\usepackage{booktabs} 
\usepackage{lscape}
\usepackage{placeins}

\usepackage{hyperref}

\begin{document}

   \title{The Nysa family as the main source of unequilibrated LL ordinary chondrites}
    \titlerunning{Nysa as the source of UOCs}
    \authorrunning{Marsset et al.}

   \author{M. Marsset\inst{1}\fnmsep\thanks{Corresponding author: mmarsset@eso.org}
        \and P. Vernazza\inst{2}
        \and M. Bro\v z\inst{3}
        \and C. Avdellidou\inst{4}
        \and C. A. Thomas\inst{5}
        \and L. McGraw\inst{5}
        \and A. Madden-Watson\inst{5}
        \and K. Minker\inst{6}
        \and M. Monnereau\inst{7}
        \and F. E.~DeMeo\inst{8}
        \and R. P.~Binzel\inst{8}
        \and M. Mahlke\inst{9}
        \and B. Carry\inst{10}
        \and J. Hanu\v s\inst{3}
        \and P. N. Simon\inst{2}
        \and B. Yang\inst{11,12}
        \and P. Beck\inst{13}
        \and M. Birlan\inst{14,15}
        \and E. Jehin\inst{16}
        }
   \institute{European Southern Observatory (ESO), Karl-Schwarzschild-Strasse 2, 85748 Garching bei München, Germany
   \and
    Aix-Marseille University, CNRS, CNES, LAM, Institut Origines, Marseille, France
\and
   Charles University, Faculty of Mathematics and Physics, Astronomical Institute, V Holešovičkách 2, CZ-18000 Praha, Czech Republic
\and
   School of Physics and Astronomy, University of Leicester, University Road, Leicester LE1 7RH, UK
\and
    Northern Arizona University, Department of Astronomy and Planetary Science, P.O. Box 6010, Flagstaff, AZ 86011, USA
\and
    Lowell Observatory, 1400 W Mars Hill Road, Flagstaff, AZ 86001, USA
\and
    IRAP, University of Toulouse, CNRS, Toulouse, France
\and
    Department of Earth, Atmospheric and Planetary Sciences, MIT, 77 Massachusetts Avenue, Cambridge, MA 02139, USA
\and
    Université Marie et Louis Pasteur, CNRS, Institut UTINAM (UMR 6213), équipe Astro, F-25000 Besançon, France
\and
    Université Côte d’Azur, Observatoire de la Côte d’Azur, CNRS, Laboratoire Lagrange, Nice, France
\and
    Instituto de Estudios Astrof\'isicos, Facultad de Ingenier\'ia y Ciencias, Universidad Diego Portales, Santiago, Chile
\and
    Planetary Science Institute, 1700 E Fort Lowell Rd STE 106, Tucson, AZ 85719, USA
\and
    Institut de Planétologie et Astrophysique de Grenoble, CNRS, Université Grenoble Alpes, 38000 Grenoble, France
\and
    Institutul Astronomic al Academiei Române, 5-Cutitul de Argint, Sector 4, 040557, Bucharest, Romania
\and
    LTE, Observatoire de Paris, 77 av Denfert Rochereau, 75014 Paris Cedex, France
\and
    Space sciences, Technologies \& Astrophysics Research (STAR) Institute, University of Liège, Liège, Belgium
}
 
  \abstract
    {The origin of the petrologic diversity observed in ordinary chondrites (OCs), the most common meteorites on Earth, remains debated. Competing models invoke either depth-dependent sampling of a single thermally stratified (“onion-shell”) parent body or contributions from multiple distinct parent bodies.}
    {We aim to determine which of the two models is preferred for LL chondrites. These are unique among OCs in exhibiting a bimodal petrologic distribution, with most meteorites being LL3 or LL6.}
    {We compare the spectral and mineralogical properties of LL chondrites and corresponding LL-chondrite-like near-Earth objects (NEOs) with their possible sources in the main asteroid belt. We also model the thermal histories of the proposed parent bodies, based on revised estimates of parent-body sizes.}
    {The spectral and mineralogical diversity of LL chondrites is consistent with contributions from the bright, S-type component of the Nysa family (Nysa\texorpdfstring{$_{\rm S}$}{S}) and the Flora family, with Nysa\texorpdfstring{$_{\rm S}$}{S} supplying mainly low-petrologic-type material and Flora higher-grade material. Unequilibrated, LL3 chondrites appear to originate exclusively from Nysa\texorpdfstring{$_{\rm S}$}{S}. Similarly, LL-chondrite-like NEOs form two distinct subpopulations consistent with origins in these same families.}
    {Our results favour multiple parent bodies for LL chondrites. The petrologic differences between the Nysa\texorpdfstring{$_{\rm S}$}{S} and Flora parent bodies can be explained by differences in their sizes, without requiring different formation times.}


   \maketitle
   \nolinenumbers

\section{Introduction}

Ordinary chondrites (OCs), the most common type of meteorites falling on our planet \citep{Gattacceca:2025}, preserve a record of the earliest stages of planetesimal formation and thermal evolution. A key distinction exists between unequilibrated OCs (UOCs; petrology type 3), which experienced peak temperatures below 800\,K \citep{Monnereau:2013} and retain pristine nebular components, and equilibrated OCs (EOCs; petrology types 4–7) that were thermally metamorphosed within their parent bodies.

UOCs consist largely of chondrules embedded in a fine-grained matrix and provide the best snapshot of the protoplanetary disk prior to substantial parent-body processing. In contrast, EOCs contain minerals that have undergone partial or complete chemical equilibration through heat-induced elemental diffusion, and chondrules become increasingly indistinct from the matrix with higher petrologic type \citep{VanSchmus:1967}.

In the canonical onion-shell model (e.g., \citealt{Minster:1979, McSween:2002, Trieloff:2003, Wood:2003, Ghosh:2006, Kessel:2007, Henke_2012A&A...545A.135H, Henke:2013, Monnereau:2013, Vernazza:2014, Gail:2019, Edwards:2020}), the distribution of petrologic types (3--7) reflects depth within a thermally stratified parent body heated primarily by the decay of ${}^{26}\mathrm{Al}$ \citep{Lee:1977}. In this framework, the relative abundance of petrologic types observed within individual compositional types of meteorites and asteroid collisional families reflects the original size and accretion timescale of the respective parent bodies.

From a mineralogical and chemical perspective, OCs are divided into three distinct groups -- H, L, and LL -- which originated from different parent bodies and differ primarily in their total iron content, the abundance of metallic iron, and the ferrous iron (FeO) content of their silicates.
The search for the parent bodies of these meteorite classes within the main asteroid belt and the near-Earth asteroid population -- primarily through comparative laboratory and astronomical spectroscopic measurements -- has been a long-standing pursuit (e.g., \citealt{Gaffey:1993, Binzel:1996, Binzel:2001, Binzel:2019, Chapman:1996, Burbine:2000, Vernazza:2008, Vernazza:2014, Thomas:2010, Popescu:2011, Reddy:2012, DeMeo:2022}).

Recent studies combining spectroscopic observations, dynamical modelling, collisional evolution, meteoritic constraints, and zodiacal dust-band associations have now successfully linked these three groups to compositionally analogous, silicate-rich (S-type; \citealt{DeMeo:2009, Mahlke:2022}) asteroid families in the main belt. In particular, H chondrites have been associated with the Karin and Koronis$_{\rm 2}$ clusters within the Koronis family \citep{Broz:2024Nature}, while L chondrites have been linked to the Massalia family \citep{Marsset:2024Nature}.

The origin of LL chondrites appears more complex. Although initially attributed to the Flora family based on spectroscopic similarities \citep{Vernazza:2008}, dynamical simulations indicate multiple source populations \citep{Broz:2024Nature}, including Flora, the S-type component of Nysa, Eunomia, and possibly Juno (which may represent a transitional composition between L and LL chondrites; \citealt{Marsset:2024Nature}). Among these, Flora and Nysa dominate the present-day flux of LL chondrites to Earth \citep{Broz:2024Nature}, consistent with their location in the inner main belt between the two most efficient delivery pathways: the $\nu_6$ secular resonance with Saturn and the 3:1 mean-motion resonance with Jupiter \citep{Granvik:2018a}. 

Interestingly, a substantial fraction of low–petrologic-type chondrites (23 out of a sample of 41) have recently been reclassified, primarily from the H and L groups to the LL group, on the basis of metal abundance determined through point counting and magnetic susceptibility measurements \citep{Eschrig:2022}.
The reclassification rate varies across datasets and is significantly higher for finds than for falls, likely reflecting the greater level of scrutiny applied to falls, which are rarer and considered more scientifically valuable (Eschrig et al., priv. comm.).

\begin{figure}[h]
\centering
\includegraphics[width=\columnwidth]{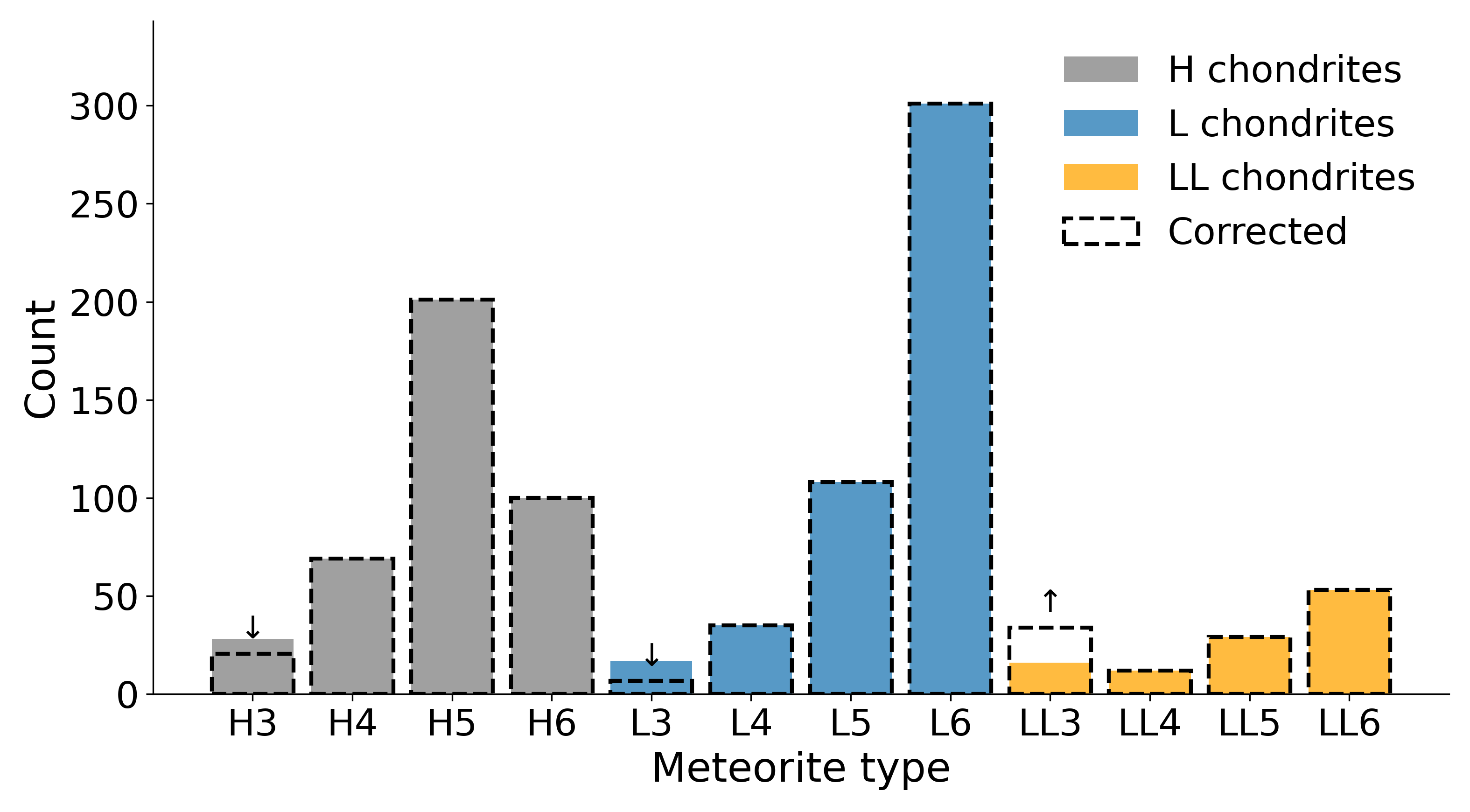}
\caption{{\bf Petrologic distribution of ordinary chondrites.} 
The filled histogram shows the distribution of meteorite falls as a function of mineralogical class (H, L, LL), compiled from the MBD. The dashed black outline shows the distribution after applying correction factors based on the revised petrologic-type ratios reported by \citet{Eschrig:2022}, as described in Appendix\,\ref{sec:app_met}. Arrows indicate that the adopted values are likely conservative, corresponding to upper ($\downarrow$) or lower ($\uparrow$) limits.}
\label{fig:histo1}
\end{figure}

As a result of this reclassification, LL chondrites now exhibit a pronounced bimodal distribution in petrologic type, with the two end members -- LL3 and LL6 -- being the most abundant (Figure\,\ref{fig:histo1}; data from the Meteoritical Bulletin Database; MBD\footnote{\url{https://www.lpi.usra.edu/meteor/}}).
Such a distribution is difficult to reconcile with the classical onion-shell model and instead suggests that LL chondrites may originate from at least two distinct source populations, responsible for the unequilibrated and equilibrated material, respectively, in agreement with recent dynamical predictions \citep{Broz:2024Nature}.

Building on these recent discoveries, we reassess the origin of LL OCs by combining asteroid and meteorite spectroscopy (Section~\ref{sec:dataset}). 
We show that S-type members of the Nysa family closely match the spectra of LL3 chondrites, whereas Flora family members resemble LL5–6 meteorites (Section~\ref{sec:petro}). 
We then demonstrate a coupled dynamical–spectroscopic correlation between LL-chondrite–like NEOs and the two asteroid families (Section~\ref{sec:dynamics}), providing direct evidence that Nysa and Flora represent the primary source populations of these objects and, therefore, LL meteorites.
Finally, we use these new associations to investigate the thermal histories of the two families’ parent bodies and discuss the implications for the survival and present-day delivery of LL3 material to Earth (Section~\ref{sec:thermal}).

\paragraph{Terminology:} The Nysa region in the inner asteroid belt -- with proper semi-major axis $2.34 \leq a_p \leq 2.48$~au, proper eccentricity $0.14 \leq e_p \leq 0.21$, 
and proper inclination $0.03 \leq \sin i_p \leq 0.06$ -- is dynamically and compositionally complex \citep{Cellino:2001, Dykhuis:2015}.  It represents a superposition of multiple overlapping collisional families, most notably the bright S-type component usually referred to as Nysa, the X-type Hertha subcluster, and the nearby C-type Polana and Eulalia families \citep{Walsh:2013}, which exhibit distinct taxonomic, albedo, and mineralogical properties. 
Asteroid (44)~Nysa itself is an interloper within the predominantly S-type family that bears its name, being classified as type~E \citep{Tedesco:2002, Gaffey:2004} in \citet{Tholen:1984}'s taxonomy and as Xn \citep{Hasegawa:2024} in the Bus-DeMeo's \citep{Bus:2002, DeMeo:2009}. For convenience, we hereafter refer to ``Nysa\texorpdfstring{$_{\rm S}$}{S}'' as the S-type component of the Nysa family.
Additional details on the historical nomenclature are provided in Appendix~\ref{sec:app_names}.

\section{Spectroscopic datasets}
\label{sec:dataset}

\begin{table*}[h!]
\begin{center}
\small
\setlength{\tabcolsep}{3pt}
\caption{Observation summary for Nysa\texorpdfstring{$_{\rm S}$}{S} family members.}\label{tab:obs}
\begin{tabular}{r l l r r l r r r r r r r}
\hline\hline
Number & Name & Designation & R.A. & Dec. & Start Time & AM & Exp. Time & N & $\Delta$ & $r$ & $\alpha$ \\
 &  &  &  &  &  &  & (s) &  & (au) & (au) & (deg) \\
\hline
11128 & Ostravia & 1996 VP &  14 36 04.0 & -12 50 24 & 2025-03-28T14:23 & 1.42 & 119.54 & 14 & 1.182 & 2.091 & 15.10 \\
29374 & Kazumitsu & 1996 GZ2 & 12 25 10.4 & -01 55 28 & 2025-03-28T12:27 & 1.30 & 119.54 & 10 & 0.959 & 1.957 & 0.70 \\
33093 &  & 1997 YF3 &  11 17 15.3 &  02 15 05 & 2025-03-28T10:15 & 1.13 & 119.54 & 20 & 1.137 & 2.108 & 8.60 \\
33372 & Jonathanchung & 1999 BP23 & 15 46 41.5 & -15 19 40 & 2025-03-28T15:17 & 1.39 & 149.66 & 10 & 1.213 & 2.004 & 22.50 \\
37823 &  & 1998 BS8 & 14 21 13.2 & -08 42 01 & 2025-03-28T13:20 & 1.25 & 149.66 & 20 & 1.193 & 2.124 & 13.00 \\
38803 &  & 2000 RH62 & 21 15 28.6 & -11 55 53 & 2025-06-23T13:04 & 1.18 & 119.54 & 18 & 1.084 & 1.935 & 22.10 \\
91215 &  & 1999 AN & 12 02 54.5 &  00 48 13 & 2025-03-28T11:22 & 1.20 & 149.66 & 20 & 1.031 & 2.025 & 3.60 \\
\hline
\end{tabular}
\end{center}

\tablefoot{The table lists the asteroid number, name, and designation; the approximate right ascension (R.A.) and declination (Dec.) at the time of observation; the start time and mean airmass (AM); the individual exposure time, and number of exposures (N); as well as the asteroid's geocentric ($\Delta$) and heliocentric ($r$) distances in astronomical units (au) and phase angle ($\alpha$) in degrees (deg) at the time of observation.}

\end{table*}

\begin{figure*}[h!]
\centering
\includegraphics[width=0.9\textwidth]{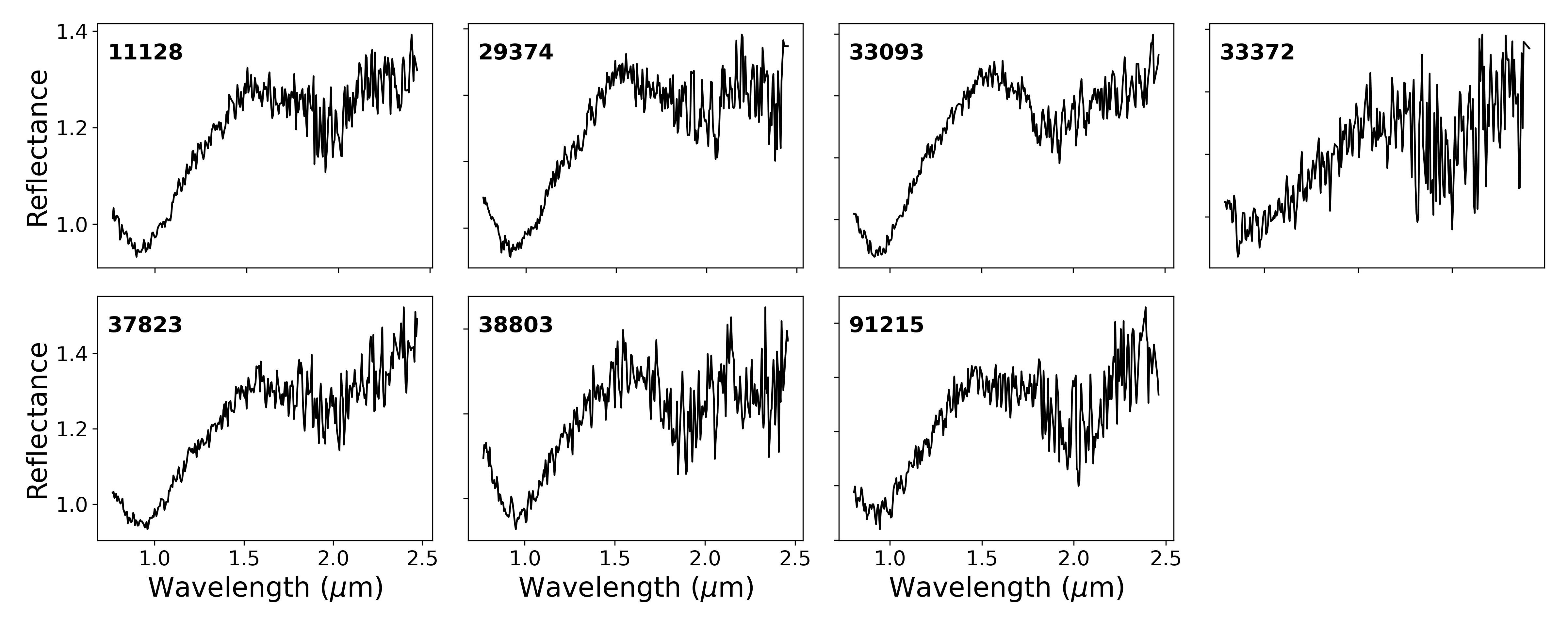}
\caption{{\bf New reflectance spectra of Nysa\texorpdfstring{$_{\rm S}$}{S} family members.}
Spectra were obtained with the SpeX spectrograph on NASA’s IRTF on 28 March and 23 June 2025 (Table\,\ref{tab:obs}).
All spectra are normalised to unity at 0.9\,$\mu$m.
The asteroid number is indicated in each panel.}
\label{fig:spec}
\end{figure*}

In this work, we reanalyse near-infrared (NIR; 0.8--2.5~$\mu$m) spectroscopic data for members of the Nysa\texorpdfstring{$_{\rm S}$}{S} (14 spectra of 13 members), Flora (47 spectra of 44 members) and Eunomia (16 spectra of 16 members) asteroid families (Appendix\,\ref{sec:app_shku}). 

Most of the data are drawn from the literature \citep{Vernazza:2008, Vernazza:2014, DeMeo:2009, Binzel:2019, Marsset:2022, Delbo:2026}. 
In addition, new reflectance spectra of seven Nysa\texorpdfstring{$_{\rm S}$}{S} family members -- selected from the family compilation of \citet{Nesvorny:2015_families} and chosen to have albedos typical of S-type asteroids (mostly $p_V \sim 0.20$–0.30; \citealt{Masiero:2011, Usui:2011, Nugent:2016, AliLagoa:2018}), in order to minimise contamination from interlopers associated with the nearby Polana, Eulalia and Hertha families -- were obtained with the SpeX spectrograph \citep{Rayner:2003} on NASA’s Infrared Telescope Facility (IRTF) on 2025 March 28 and 2025 June 23 (Program ID 2025A016; PI: Marsset). 
Observing circumstances are summarised in Table~\ref{tab:obs}.

These asteroid observations were accompanied by measurements of the solar-analogue stars L107-684, L107-998, L110-361, and L113-276 \citep{Landolt:1973}, which are widely used for spectroscopic calibration in asteroid surveys \citep{Marsset:2020_Spex}. 

Data reduction and spectral extraction followed the same procedure as used in the MITHNEOS survey, as described by \citet{Binzel:2019}, which we summarise here.
The two-dimensional spectral images were reduced using the Image Reduction and Analysis Facility (IRAF) and Interactive Data Language (IDL), with the Autospex software tool \citep{Cushing:2004} used to generate automated command files. 

Reduction steps for the science targets and their associated calibration stars included image trimming, bad-pixel masking, flat-field correction, sky subtraction using AB image pairs, spectral tracing in both the spatial and wavelength directions, co-addition of individual exposures, spectral extraction, wavelength calibration, and telluric correction with the ATRAN software \citep{Lord:1992} to consider airmass differences between the asteroids and their corresponding solar analogues. 
The asteroid spectra were divided by the mean stellar spectra to remove the solar continuum. 

The resulting reduced reflectance spectra are shown in Figure\,\ref{fig:spec}.
Classification within the Bus–DeMeo taxonomy \citep{Bus:2002, DeMeo:2009}, based on mean-squared-error comparisons with average class spectra implemented in the M4AST tool \citep{Popescu:2012}, yields Sq-types for (11128) Ostravia, (33372) Jonathanchung, and (37823) 1998~BS8, and S-types for (29374) Kazumitsu, (33093) 1997~YF3, (38803) 2000~RH62, and (91215) 1999~AN.

We also make use of 133 laboratory spectra of 70 unique LL ordinary chondrites from the RELAB (99 spectra; \citealt{Pieters:2004}) and SSHADE \citep{Schmitt:2018} databases, more specifically the dataset of \citet{Eschrig:2022} (34 spectra). For the latter, we consider the petrologic classification of \citet{Bonal:2016} and the revised H/L/LL compositional classification of
\citet{Eschrig:2022}. The list of meteorite spectra used in this study is provided in Appendix~\ref{sec:app_met}. 

Finally, we consider the largest available NIR (0.8--2.5~$\mu$m) spectroscopic dataset of NEOs from the MITHNEOS survey \citep{Binzel:2019, Marsset:2022}, focusing on the 325 S-type NEOs spectrally analogous to LL chondrites as identified by \citet{Marsset:2024Nature} (395 spectra). The list is provided in Appendix~\ref{sec:app_neo}.

\section{Petrologic types of the Nysa\texorpdfstring{$_{\rm S}$}{S} and Flora parent bodies}
\label{sec:petro}

\subsection{Spectral comparison}
\label{sec:spectra}

\begin{figure}[h!]
\centering
\includegraphics[width=0.9\columnwidth]{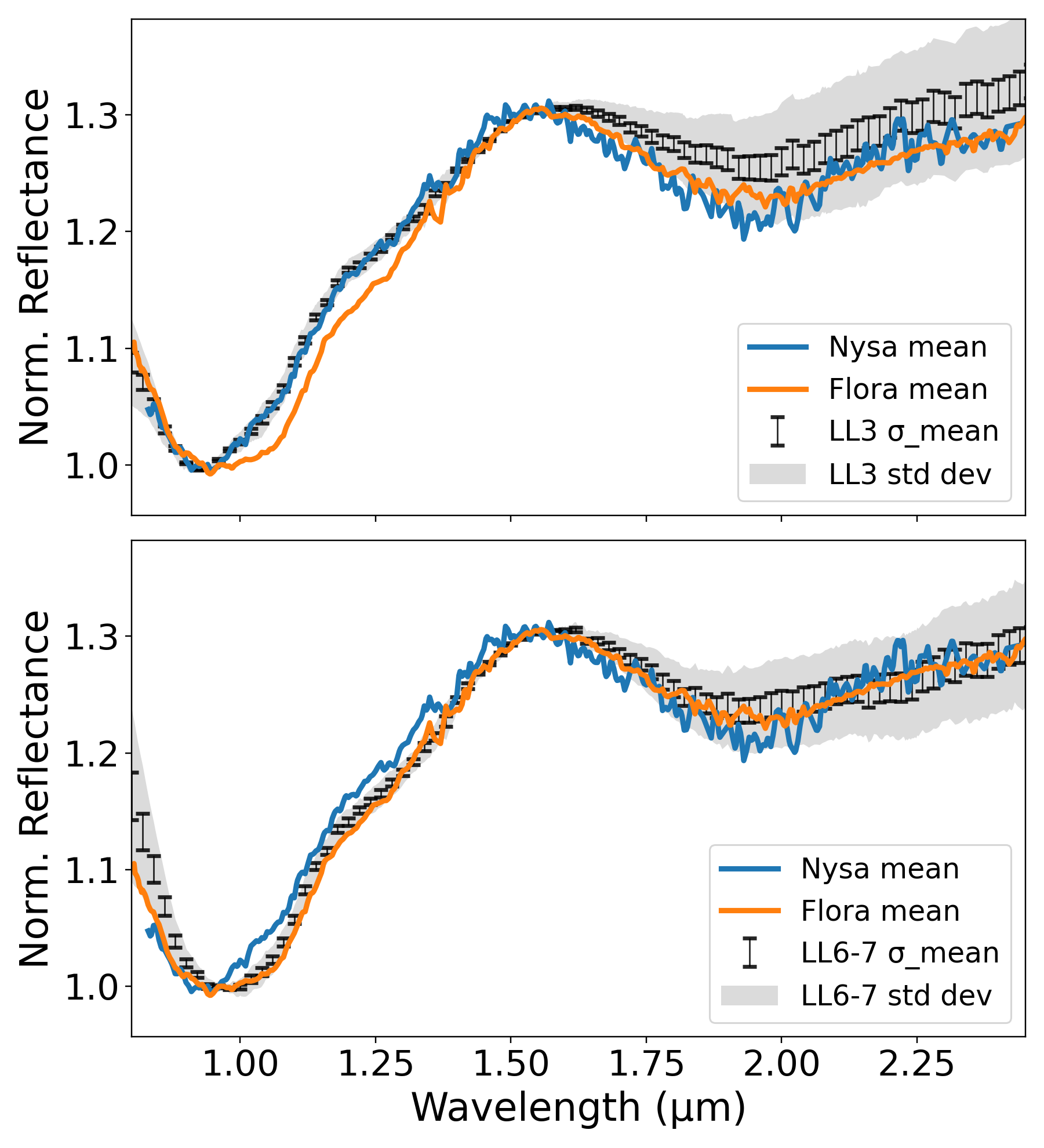}
\caption{{\bf Spectral comparison between LL ordinary chondrites and the Nysa\texorpdfstring{$_{\rm S}$}{S} and Flora families.} 
Average near-infrared reflectance spectra of unequilibrated (petrologic type~3; top) and highly equilibrated (petrologic types~6 and~7; bottom) LL chondrites are compared with the average spectra of the Nysa\texorpdfstring{$_{\rm S}$}{S} (blue) and Flora (orange) asteroid families.
All data are normalised near the 1~$\mu$m band minimum and dereddened using a space-weathering exponential model (see Section\,\ref{sec:spectra}). The small discontinuity near 1.9~$\mu$m observed in the average LL3 spectrum is caused by the detector transition of the SHADOWS spectrometer \citep{Potin:2018} used to acquire the SSHADE dataset. 
The comparisons show that unequilibrated LL3 meteorites are highly consistent with the Nysa\texorpdfstring{$_{\rm S}$}{S} family, whereas equilibrated LL6–7 meteorites preferentially match Flora, particularly in the shape of the 1~$\mu$m absorption band -- the most diagnostic feature of silicate composition and, in particular, of the olivine-to-pyroxene ratio. This dual correspondence supports a multiple-parent-body origin for LL chondrites rather than a single onion-shell structure.}
\label{fig:meteorites}
\end{figure}

\begin{figure*}[h!]
\centering
\includegraphics[width=0.8\textwidth]{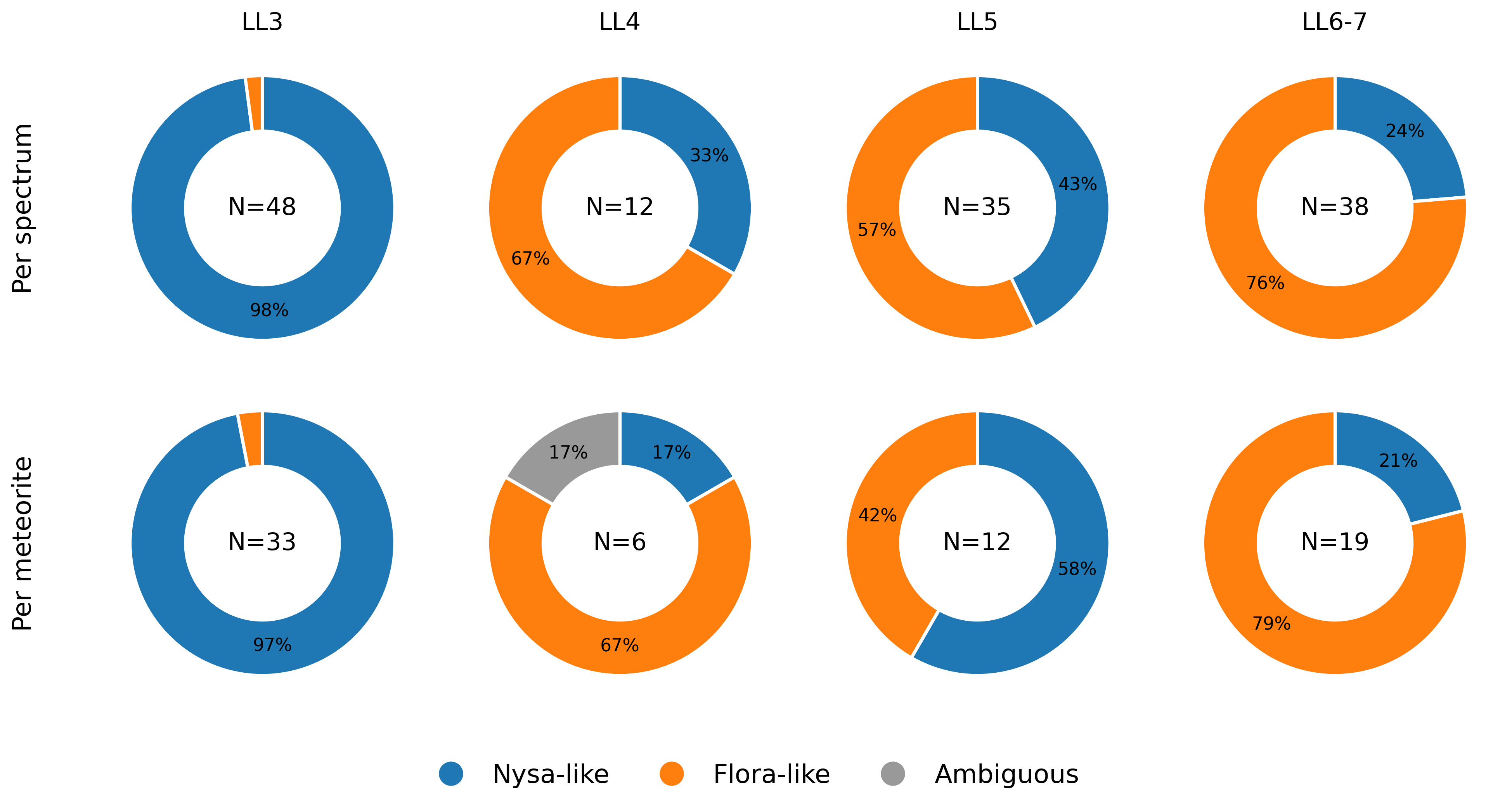}
\caption{{\bf Fractions of LL chondrites spectrally associated with the Nysa\texorpdfstring{$_{\rm S}$}{S} and Flora families.} 
Pie charts show, for each petrologic type, the fraction of meteorite spectra (top) and meteorites (bottom) assigned to each family based on the classifications in Appendix~\ref{sec:app_comp}. The numbers $N$ indicate the total number of spectra or unique meteorites in each subtype.  
LL3 meteorites are almost exclusively associated with Nysa\texorpdfstring{$_{\rm S}$}{S}-like spectra, whereas the more equilibrated LL chondrites (LL4--7) are dominated by Flora-like spectra. One meteorite, LL4 Hamlet, has two spectra best matched to Flora and two to Nysa\texorpdfstring{$_{\rm S}$}{S}; we classify it as ambiguous.}
\label{fig:pies}
\end{figure*}

\begin{figure}[h!]
\centering
\includegraphics[trim={0 0 0 0}, clip, width=0.9\columnwidth]{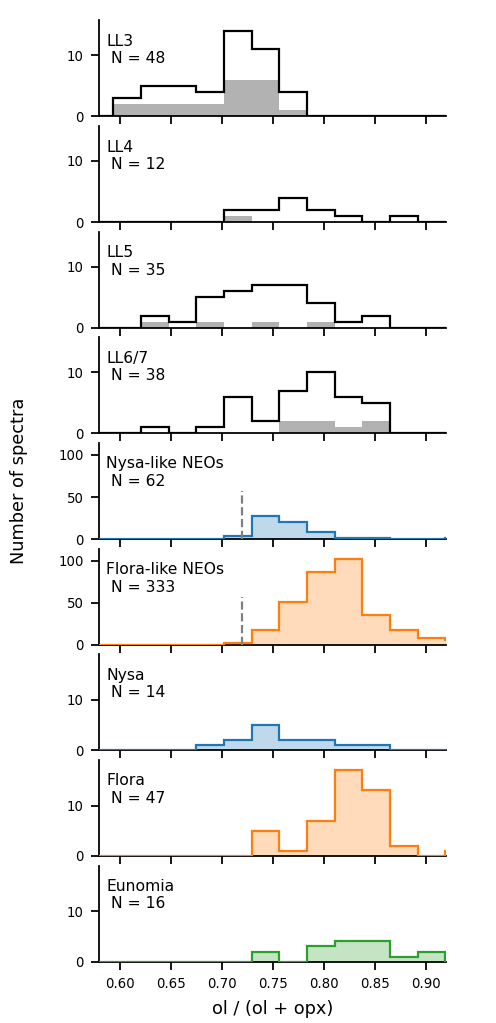}
\caption{{\bf Mineralogy of LL chondrites, LL-chondrite-like NEOs
and corresponding asteroid families.} 
Histograms of the ol/(ol\,+\,opx) ratios derived for each spectrum from the Shkuratov model show that the Nysa\texorpdfstring{$_{\rm S}$}{S} family exhibits values close to those of unequilibrated, type-3 LL chondrites, whereas the Flora and Eunomia families are shifted toward values characteristic of more thermally equilibrated, higher-petrologic-type material.
Approximately 3\% offsets between LL3 and Nysa\texorpdfstring{$_{\rm S}$}{S}, and between LL6 and Flora, are likely due to limitations of the model, which is sensitive to differences in space-weathering states between meteorites and asteroids.
In the meteorite panels, empty black histograms correspond to the combined RELAB+SSHADE dataset, while filled grey histograms show SSHADE \citep{Eschrig:2022} data only.
For the NEO panels, the sample is restricted to objects with a mean ol/(ol+opx) ratio ${>} 0.72$, corresponding to the dashed vertical line, in order to minimise contamination from L-chondrite-like NEOs associated with the Massalia family \citep{Marsset:2024Nature}. 
}
\label{fig:histo2}
\end{figure}

Although the Flora, Nysa\texorpdfstring{$_{\rm S}$}{S}, Eunomia, and Juno (L/LL) families are all classified as LL-chondrite-like, they exhibit subtle spectroscopic differences (Appendix\,\ref{sec:app_shku}) that allow their respective meteorite analogues to be identified within laboratory collections \citep{Marsset:2024Nature}.  

Guided by dynamical models indicating that the Flora and Nysa\texorpdfstring{$_{\rm S}$}{S} families dominate the present-day flux of LL chondrites to Earth \citep{Broz:2024Nature}, we therefore focus on these two families and assume that contributions from other LL-like families (Eunomia and Juno), as well as from the background main-belt population, are negligible.  
Specifically, \citet{Broz:2024Nature} predict that, at metre scale, Flora contributes $\sim$73\% of LL chondrites reaching Earth and Nysa\texorpdfstring{$_{\rm S}$}{S} $\sim$16\%. At kilometre scale, Flora contributes $\sim$72\% of LL-like NEOs and Nysa\texorpdfstring{$_{\rm S}$}{S} $\sim$9\%.

Family-average spectra are constructed by weight-averaging individual member spectra after removing the overall spectral slope using the empirical exponential function of \citet{Brunetto:2006} to account for space-weathering effects \citep{Pieters:2000, Sasaki:2001, Chapman:2004}. 
We adopt the average LL chondrite spectrum of \citet{Vernazza:2014} as the reference template for unweathered material. 

Specifically, the dereddening procedure is performed such that the maximum reflectance of each spectrum -- located near 1.6\,$\mu$m, between the two silicate absorption bands -- coincides with the maximum of the LL-chondrite reference spectrum, after normalisation at the wavelength corresponding to the minimum of the 1\,$\mu$m silicate absorption band. This approach facilitates a direct comparison of the 1\,$\mu$m band, the spectral feature most sensitive to silicate composition, and in particular to the olivine-to-pyroxene ratio.

We then assess whether the distinct spectroscopic properties of the Nysa\texorpdfstring{$_{\rm S}$}{S} and Flora families can be attributed to differences in petrologic type by performing direct $\chi^2$ comparisons between their average spectra and laboratory spectra of LL chondrites from the RELAB and SSHADE databases. 
Using the petrographic metadata, the meteorites were separated into unequilibrated LL3.x and highly equilibrated LL6-7.x subtypes, and an average composite spectrum was constructed for each group. 
Only spectra falling within a $\chi^2$ threshold
relative to at least one of the two family averages were retained.
A fraction of unfitted spectra correspond to measurements performed on meteorite chips and slabs (as opposed to size-sorted, ground particles), which often exhibit spectral slopes and band depths that differ from those of asteroid regolith.
See Appendix\,\ref{sec:app_comp} for further details on the construction of the average meteorite spectra, the $\chi^2$ computation, and the sensitivity of meteorite spectra to sample texture and preparation.

Figure~\ref{fig:meteorites} shows the resulting spectral comparison between LL chondrites and the Nysa\texorpdfstring{$_{\rm S}$}{S} and Flora families.
Reflectance spectra of unequilibrated LL3 meteorites exhibit an excellent match to the Nysa\texorpdfstring{$_{\rm S}$}{S} family, whereas equilibrated LL6--7 meteorites are instead highly similar to the Flora family. 
Appendix~\ref{sec:app_comp} presents additional examples of high-quality spectral fits for individual meteorites with both families.

Moreover, when we force each of the 133 individual meteorite spectra to be assigned to either Nysa\texorpdfstring{$_{\rm S}$}{S} or Flora based on the $\chi^2$ metric -- independently of the absolute goodness of fit -- we find that all but one LL3 meteorites are associated with Nysa\texorpdfstring{$_{\rm S}$}{S}, while higher petrologic types preferentially match Flora, with some additional association with Nysa\texorpdfstring{$_{\rm S}$}{S} (Figure~\ref{fig:pies}). 

While a clear trend emerges from these spectral associations, individual matches should be interpreted with caution, as spectral similarity does not necessarily imply a direct genetic link. This is particularly true for unequilibrated and weakly equilibrated OCs, which can exhibit significant variability in mineralogical parameters such as Mg/(Mg+Fe) (Mg\#) and ol/(ol+opx). In addition, laboratory spectra are measured on small samples and may not fully represent bulk compositions, while both asteroid families and meteorite groups exhibit intrinsic compositional dispersion. In this context, the apparent association of some LL5–6 meteorites with Nysa\texorpdfstring{$_{\rm S}$}{S} (Figure~\ref{fig:pies}) -- which is unexpected from a thermal evolution perspective (see Section~\ref{sec:thermal}) -- should be treated with caution.

\subsection{Radiative-transfer model}
\label{sec:shku}

The H, L, and LL mineralogical groups of OCs exhibit progressively higher olivine-to-low-calcium pyroxene (orthopyroxene) ratios from H to L and LL compositions. 
This mineralogical trend produces distinct spectral signatures at optical and NIR wavelengths, enabling robust classification of OC-like bodies through compositional modelling using radiative transfer models (e.g., \citealt{Hapke:1981, Shkuratov:1999}). 
Subtle differences in silicate composition also exist among LL-chondrite--like asteroid families \citep{Vernazza:2014, Marsset:2024Nature}, allowing the compositional affinities of meteorites to be investigated using the same approach.

Here, we model the silicate compositions of Flora and Nysa\texorpdfstring{$_{\rm S}$}{S} family members using the radiative transfer model of \citet{Shkuratov:1999}, which simulates light scattering in mixed mineral assemblages. 
Specifically, we use the IDL implementation of the model developed by \citet{Vernazza:2008, Vernazza:2014}.\footnote{available at \url{https://github.com/mmarsset/Shkuratov-IDL/}}
For a direct, one-to-one comparison with LL chondrites, we apply the same analysis to our compiled dataset of laboratory meteorite spectra.

Our model employs olivine (ol), orthopyroxene (opx), and chromite as end-member components calibrated with laboratory meteorite spectra. 
Space-weathering effects are included through an empirical exponential reddening function \citep{Brunetto:2006}, and spectral fitting is restricted to wavelengths below 1.9~$\mu$m to avoid known model overestimation at longer wavelengths. 
Best-fit mineral abundances are derived via least-squares minimisation, with the olivine fraction, chromite fraction, grain size, and space-weathering coefficient treated as free parameters. The Fe/Mg ratio of the silicates is fixed at 70/30.

The results are shown in Figure~\ref{fig:histo2} and the corresponding model parameters are provided in Appendix~\ref{sec:app_shku}. 
Overall, the modelled olivine-to-pyroxene ratio, ol/(ol + opx), increases systematically with petrologic type in the meteorite dataset, from 0.71 for LL3 to 0.76 for LL4, 0.75 for LL5, and 0.79 for LL6/7 (median values).
Most Nysa\texorpdfstring{$_{\rm S}$}{S} family members fall within the range of values characteristic of LL3 chondrites, whereas Flora family members fall mostly in the range of LL5-7 chondrites.
In addition, the offset between the median values of the Nysa\texorpdfstring{$_{\rm S}$}{S} and Flora distributions is 8.3\%, which is remarkably similar to the 7.7\% offset observed between LL3 and LL6 chondrites, strongly suggesting that these two families represent the end-member petrologic sources of LL chondrites.

About 3\% offsets between the median ratios of LL3 and Nysa\texorpdfstring{$_{\rm S}$}{S}, and between LL6 and Flora are likely due to limitations of the model, which is sensitive to differences in space-weathering states between meteorites and asteroids. 
These effects on spectral shape may be more complex than the simple exponential function adopted in our implementation of the Shkuratov model.

Overall, the derived mineralogies support a shared origin for LL3 meteorites and the Nysa\texorpdfstring{$_{\rm S}$}{S} family, and between higher petrologic types and the Flora family, although Nysa\texorpdfstring{$_{\rm S}$}{S} may have contributed to a fraction of the more equilibrated material.

The progression in bulk silicate mineralogy observed among LL chondrites with increasing petrologic type may reflect 
(1) an increasing contribution from the Flora family and a decreasing contribution from the Nysa\texorpdfstring{$_{\rm S}$}{S} family to the LL meteorite record, 
and/or 
(2) chemical modification driven by parent-body thermal metamorphism. Progressive equilibration during metamorphism involves oxidation and redistribution of iron into silicate phases, which has been shown to favor higher olivine abundances at the expense of metallic Fe and low-Ca pyroxene, for example through the reaction:
\begin{equation}
\mathrm{Fe + O + (Mg,Fe)SiO_3 \rightarrow (Mg,Fe)_2SiO_4},
\end{equation}
where the oxygen may be supplied by trace amounts of ice (ice-to-rock ratios at the per mil level; \citealt{McSween:1991,McSween:1993}).

\section{Dynamical arguments}
\label{sec:dynamics}

\subsection{Orbital distributions of NEOs}
\label{sec:neos}

NEOs originating from a given asteroid family retain dynamical signatures of their source region, particularly in the distribution of semi-major axis $a$ versus inclination $i$ \citep{Broz:2024, Broz:2024Nature, Marsset:2024Nature}. In addition, as shown above for the meteorite dataset, individual petrologic types can be distinguished using reflectance spectroscopy, despite some overlap between classes. Demonstrating a combined spectroscopic–orbital correlation between NEOs and asteroid families would therefore provide strong additional evidence for the existence of multiple source populations.

Here we consider spectroscopic data of NEOs from the MITHNEOS survey \citep{Binzel:2019,Marsset:2022}, focusing on S-type NEOs spectrally analogous to LL chondrites, as identified by \citet{Marsset:2024Nature}. To minimise contamination from L-chondrite-like objects associated with the Massalia family, we exclude bodies with $\mathrm{ol}/(\mathrm{ol}+\mathrm{opx}) < 0.72$. This threshold is motivated by the fact that all studied Massalia family members exhibit $\mathrm{ol}/(\mathrm{ol}+\mathrm{opx}) < 0.70$ \citep{Marsset:2024Nature}; adopting a slightly higher cutoff provides a margin to account for the limited sampling of that family and possible variability in its silicate compositions.
For objects with multiple spectra, we adopt the object-averaged $\mathrm{ol}/(\mathrm{ol}+\mathrm{opx})$ value. Individual spectra with $\mathrm{ol}/(\mathrm{ol}+\mathrm{opx}) < 0.72$ may therefore be retained if the mean value for the object exceeds the threshold. The resulting dataset, which comprises 395 spectra of 325 individual NEOs, is presented in Appendix\,\ref{sec:app_neo}.

Each object is classified as Nysa\texorpdfstring{$_{\rm S}$}{S}-like or Flora-like through direct $\chi^2$ comparisons of its reflectance spectrum with the corresponding family-average spectra, following the same methodology applied to the meteorite data.
Prior to classification, the NEO spectra are dereddened using the same procedure as for the family members.
This analysis identifies 53 Nysa\texorpdfstring{$_{\rm S}$}{S}-like bodies, 269 Flora-like bodies, and 3 ambiguous cases, for which multiple spectroscopic measurements yield inconsistent classifications (out of 43 bodies with multiple measurements available).
Bodies with ambiguous family assignments include (1943)~Anteros, (6455)~1992~HE, and (22753)~1998~WT.

The resulting ratio of Flora-like to Nysa\texorpdfstring{$_{\rm S}$}{S}-like NEOs in our sample is 5.1, for objects with sizes predominantly in the range $\sim$100\,m–2\,km. This value is in good agreement with the dynamical predictions of \citet{Broz:2024Nature}, who find Flora-to-Nysa\texorpdfstring{$_{\rm S}$}{S} ratios of $\sim$4.6 at metre scale and $\sim$8.0 at kilometre scale in the NEO population.

\begin{figure}
\centering
\includegraphics[width=0.95\columnwidth]{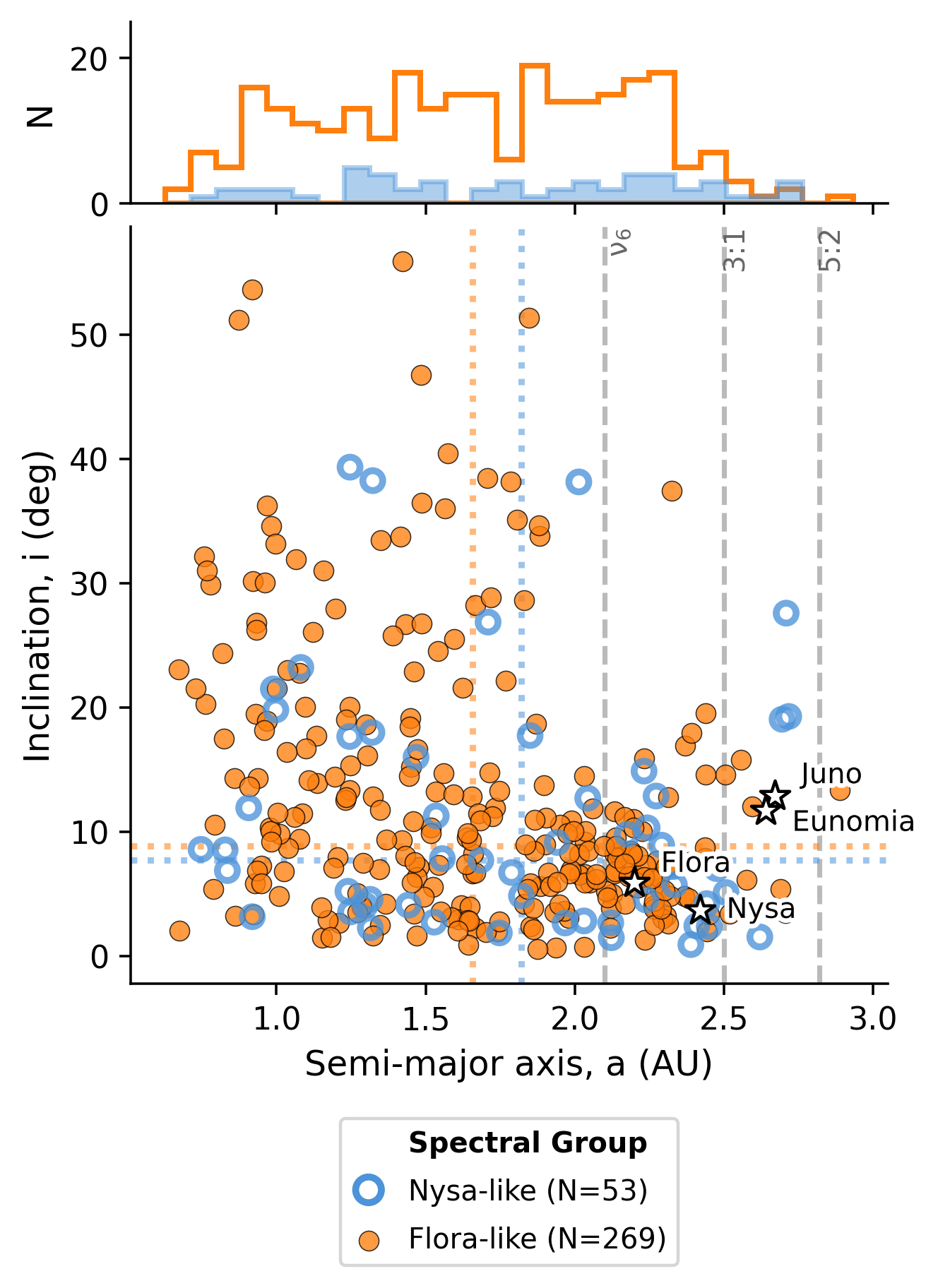}
\caption{{\bf Orbital distribution of LL-chondrite-like NEOs in semi-major axis versus inclination.} The top panel shows the semi-major-axis distribution of the same objects (histograms), using the same colour coding as in the scatter plot. Colours indicate the best spectral family analogue (blue: Nysa\texorpdfstring{$_{\rm S}$}{S}, orange: Flora). Three objects with inconsistent classifications across multiple spectral measurements are not shown. Dotted lines show the median values of $a$ and $i$ for Nysa\texorpdfstring{$_{\rm S}$}{S}-like and Flora-like NEOs. Black stars mark the orbital locations of the Nysa\texorpdfstring{$_{\rm S}$}{S}, Flora, Juno, and Eunomia families in the main belt. Vertical dashed lines indicate the $\nu_6$ secular resonance with Saturn and the 3:1 and 5:2 mean-motion resonances with Jupiter. The offset between the Nysa\texorpdfstring{$_{\rm S}$}{S} and Flora families in the main belt ($\Delta_a$=0.22~au, $\Delta_i$=2.2$\degr$) appears imprinted in the NEO population ($\Delta \tilde{a} = 0.18$~au, $\Delta \tilde{i} = 1.1^\circ$), supporting these families as the principal sources of LL-chondrite–like NEOs and, consequently, LL meteorites (see Section\,\ref{sec:neos}). Very few NEOs are located near the Juno and Eunomia families, indicating that these families contribute only marginally to the NEO population.}
\label{fig:orbits}
\end{figure}

The resulting mineralogical distribution of the classified NEOs is shown in Figure~\ref{fig:histo2}.
Figure~\ref{fig:orbits} shows the distribution of semi-major axis $a$ versus inclination $i$ for the Nysa\texorpdfstring{$_{\rm S}$}{S}-like and Flora-like NEO populations. The median orbital elements of the two groups differ by $\sim$0.18~au in $a$ and $\sim$1.1$^\circ$ in $i$. These offsets are qualitatively consistent with the separation between the Flora and Nysa\texorpdfstring{$_{\rm S}$}{S} families in the main belt ($\Delta a = 0.22$~au, $\Delta i = 2.2^\circ$), although they are expected to be reduced by dynamical excitation and chaotic diffusion during transport into near-Earth space. In particular, the offset in semi-major axis may reflect a larger contribution of Nysa\texorpdfstring{$_{\rm S}$}{S}-like NEOs delivered through the 3:1 mean-motion resonance, consistent with the location of the family in the main belt.

Bootstrapping $10^{6}$ times with replacement, we assess the significance of the observed separation between the Nysa\texorpdfstring{$_{\rm S}$}{S}-like and Flora-like NEO populations in $(a,i)$ space.
In each realization, two random subsamples are drawn from the pooled sample of LL-chondrite-like NEOs (including Nysa\texorpdfstring{$_{\rm S}$}{S}-like, Flora-like, and ambiguous objects), with sizes fixed to match the observed numbers of Nysa\texorpdfstring{$_{\rm S}$}{S}-like and Flora-like objects.
For each bootstrap realization, we compute the differences in median orbital elements,
$\Delta a = \mathrm{median}(a_{\rm A}) - \mathrm{median}(a_{\rm B})$ and
$\Delta i = \mathrm{median}(i_{\rm A}) - \mathrm{median}(i_{\rm B})$,
and compare them to the observed offsets between the Nysa\texorpdfstring{$_{\rm S}$}{S}-like and Flora-like populations.

We find that only 3.3\% of the bootstrap realizations yield a separation at least as extreme as observed, in the sense of simultaneously having
$\Delta a \ge \Delta a_{\rm obs}$ and $\Delta i \le \Delta i_{\rm obs}$ (here, $\Delta i_{\rm obs}$ is negative).
This low joint-tail probability indicates that the observed offset -- where the Nysa\texorpdfstring{$_{\rm S}$}{S}-like population occupies a systematically higher-$a$, lower-$i$ region of orbital-element space than the Flora-like population -- is unlikely to arise by chance from the overall LL-like NEO distribution.
These results support the interpretation that the two NEO populations originate predominantly from distinct parent families associated with Flora and Nysa\texorpdfstring{$_{\rm S}$}{S}.

Very few NEOs are found near the orbital loci of the other LL-chondrite-like families, Juno and Eunomia (Figure\,\ref{fig:orbits}), indicating that their contribution to the observed NEO population is marginal. In addition, any residual misclassification involving these families would tend to increase the mean inclination of the Nysa\texorpdfstring{$_{\rm S}$}{S}-like group more than that of the Flora-like group, because (i) Juno, which contributes more NEOs than Eunomia \citep{Broz:2024Nature}, is spectrally closer to Nysa\texorpdfstring{$_{\rm S}$}{S}, whereas Eunomia is spectrally closer to Flora \citep{Marsset:2024Nature}, and (ii) Flora-like NEOs are more numerous than Nysa\texorpdfstring{$_{\rm S}$}{S}-like NEOs, making their mean inclination less sensitive to contamination by a relatively small number of high-inclination interlopers. This acts in the opposite direction than the observed trend, therefore reinforcing our conclusions.

\subsection{Pre-atmospheric orbits of meteorites}
\label{sec:pre-atm}

\def\s{\phantom{0}}

\begin{table}[h!]
\centering
\caption{
Orbital elements for LL meteorite falls and asteroid Itokawa,
with probabilities of origin (in \%) from the Flora and Nysa\texorpdfstring{$_{\rm S}$}{S} asteroid families.
}
\small
\begin{tabular}{@{}l@{\ \ }l@{\ \ }c@{\ \ }c@{\ \ }c@{\ }c@{\ }c@{}}
\hline\hline
Meteorite & Type & $a$ & $e$ & $i$ & $P$ [Flora] & $P$ [Nysa\texorpdfstring{$_{\rm S}$}{S}] \\
 &  & (AU) & & ($^\circ$) & (\%) & (\%) \\
\hline
Bene\v{s}ov$^1$     & LL3.5, H5 & 2.483 & 0.627 & 24.0   & 73.8 &   26.2 \\
Chelyabinsk         & LL5           & 1.760 & 0.580 & \s 4.9 & 87.9 &   12.1 \\
Dingle Dell         & L/LL6         & 2.254 & 0.590 & \s 4.1 & 82.2 &   17.8 \\
Dishchii’bikoh$^2$  & LL7           & 1.129 & 0.206 & 21.2   & 99.0 & \s 1.0 \\
Haag                & LL4-6         & 1.844 & 0.561 & \s 2.9 & 88.9 &   11.1 \\
Hradec Králové      & LL5           & 1.721 & 0.413 & \s 8.2 & 90.0 &   10.0 \\
Ischgl              & LL6           & 1.223 & 0.258 & 31.6   & 88.8 &   11.2 \\
Stubenberg          & LL6           & 1.525 & 0.394 & \s 2.1 & 76.9 &   23.1 \\
\\
Asteroid & & & & & & \\
\hline
Itokawa            & LL5       & 1.324 & 0.280 & \s 1.6 & 93.1 & \s 6.9 \\
\\
\hline
\end{tabular}
\label{tab:pre-atm}
\tablefoot{
$^1$Polymict breccia \citep{Spurny_2014A&A...570A..39S},
alternatively from Eunomia (59.4\%) in a four-families model;
$^2$Alternatively from Juno (57.4\%) in a four-families model.
}
\end{table}

\begin{table}[h!]
\caption{Parent body sizes (in km) of OC families.}
\label{tab:pb_sizes}
\centering
\small
\begin{tabular}{lr@{\ \ }r@{\ \ }r@{\ \ }r@{\ $\rightarrow$\ }r@{\ $\rightarrow$\ }r@{\ $\rightarrow$\ }r}
\hline\hline
\vline width 0pt depth 4pt
Family              & $q_1$  & $q_2$  & $D_0$ & obs.  & 1\,km & 1\,m    & $100\,\mu{\rm m}$ \\
\hline
\vline height 10pt width 0pt
Koronis         & $-2.8$ &        &       &  99   & 182   & 215     & 223       \\
Koronis         & $-2.9$ &        &       &  99   & 182   & 230     & {\bf 253} \\
Koronis         & $-3.0$ &        &       &  99   & 182   & 254     & 319       \\
Koronis$_2$     & $-4.0$ & $-3.0$ & 0.2   &  34.6 & 34.7  & 43.6    & 62.8      \\
Karin           & $-3.0$ &        &       &  18.5 & 20.5  & 31.8    & 41.5      \\
\multicolumn{7}{l}{H chondrites${}^{1}$}                            & 240--280  \\
\\
Massalia        & $-5.7$ & $-2.8$ & 0.5   & 132   & 132   & 143     & 145       \\
Massalia        & $-5.7$ & $-3.0$ & 0.5   & 134   & 134   & 151     & {\bf 171} \\
Massalia        & $-5.7$ & $-3.2$ & 0.5   & 154   & 154   & 179     & 294       \\
Massalia$_2$    & $-2.8$ &        &       & 131   & 131   & 132     & 132       \\
Massalia$_2$    & $-3.0$ &        &       & 131   & 131   & 133     & 134       \\
Massalia$_2$    & $-3.2$ &        &       & 131   & 131   & 134     & 154       \\
\multicolumn{7}{l}{L chondrites${}^{2}$}                            & 200--320  \\
\\ 
Flora           & $-2.9$ &        &       & 153   & 158   & 169     & 176       \\
Flora           & $-3.0$ &        &       & 153   & 158   & 175     & {\bf 196} \\
Flora           & $-3.1$ &        &       & 153   & 158   & 184     & 246       \\
\multicolumn{7}{l}{LL4-6 chondrites${}^{3}$}                        & $>$300    \\
\\
Nysa\texorpdfstring{$_{\rm S}$}{S}  & $-5.3$ & $-2.7$ & 1.5   &  33   & 52    & 73      & 75        \\
Nysa\texorpdfstring{$_{\rm S}$}{S}  & $-5.3$ & $-2.7$ & 1     &  33   & 56    & 87      & {\bf 90}  \\
Nysa\texorpdfstring{$_{\rm S}$}{S}  & $-5.3$ & $-2.7$ & 0.5   &  33   & 56    & 118     & 122       \\
\multicolumn{7}{l}{LL3 chondrites} & \\
\\
Eunomia         & $-5.0$ & $-2.7$ & 5     & 272   & 331   & 366     & 371       \\ 
Eunomia         & $-5.0$ & $-2.7$ & 4     & 272   & 342   & 385     & {\bf 390} \\ 
Eunomia         & $-5.0$ & $-2.7$ & 3     & 272   & 356   & 412     & \s{419}   \\ 
\hline
\end{tabular}

\tablefoot{
Parent body sizes were estimated by power-law extrapolations of their family's SFD, $N = CD^{q}$, constrained by the currently observed SFD \citep{Broz:2024Nature,Marsset:2024Nature}. 
The SFD extrapolation down to a given size is denoted by the $\rightarrow$ symbol.
For each family, we provide the volume-equivalent diameter of the parent body obtained by integrating the volume implied by the SFD over the observed size range (“obs.”), and extrapolated down to 1\,km, 1\,m, and 100\,$\mu$m. For the extrapolation, we adopt either a single power law ($q_1$) or a broken power law ($q_1$, $q_2$) with a transition at diameter $D_0$ (in km).
The preferred parent body sizes (bold) correspond
to the intermediate SFD slopes and extrapolations
down to 100\,$\mu$m.
Methods based on  SPH simulations \citep{Durda:2007} do not have sufficient resolution (do not produce any meteoroids or dust), which leads to underestimated sizes and to results more-or-less equivalent to extrapolations truncated at $\sim$1~km.
For comparison, we provide parent-body size estimates derived from thermal models calibrated using thermometric data:
${}^{1}$\cite{Henke_2012A&A...545A.135H, Monnereau:2013, Blackburn:2017},
${}^{2}$\cite{Blackburn:2017, Gail:2019},
${}^{3}$\cite{Edwards:2020}.}
\end{table}

Large-scale camera networks (e.g., \citealt{Jenniskens:2011, Bland:2012, Borovicka:2019, Borovicka:2022, Colas:2020, Vida:2021, Toth:2026}) record atmospheric entries and enable the reconstruction of pre-atmospheric orbits of meteors, thereby providing a direct link between recovered samples and their source regions in the main asteroid belt \citep{Granvik:2018b, Jenniskens:2025}.
Reliable orbital solutions are currently available for eight LL chondrites listed in Table~\ref{tab:pre-atm}, including one unequilibrated meteorite (Bene\v{s}ov).

We used their orbital parameters as input to the METEOMOD orbital distribution model for metre-sized objects \citep{Broz:2024} to estimate the probability that each meteorite originated from either the Nysa\texorpdfstring{$_{\rm S}$}{S} or Flora asteroid families (as if no other source populations are considered). Even though individual meteorites show a preferential association with Flora, it does not necessarily imply that all of them originated from this family. From a dynamical perspective, a certain fraction of LL meteorites are expected to originate from Nysa\texorpdfstring{$_{\rm S}$}{S} \citep{Broz:2024Nature}. Notably, the only unequilibrated meteorite in the sample, Bene\v{s}ov, has the highest probability of originating from Nysa\texorpdfstring{$_{\rm S}$}{S} (26.2\%).

Considering all four LL-like families -- Flora, Nysa\texorpdfstring{$_{\rm S}$}{S}, Eunomia, and Juno -- we find Eunomia to be the most likely dynamical source of Bene\v{s}ov, with a probability of 59.4\%, as noted by \citet{Jenniskens:2025}.
However, we consider this association unlikely on compositional grounds.
The mineralogy of the Eunomia family lies at the high end of the ol/(ol+opx) distribution of LL chondrites (Figure~\ref{fig:histo2}), indicating a parent body dominated by highly equilibrated material, even more so than Flora.

On the other hand, equilibrated LL chondrites are still preferentially associated with Flora, based on both their pre-atmospheric orbits and their modelled mineralogies.
The sole exception is the LL7 Dishchii’bikoh, which seems dynamically related to Juno -- an association that is itself unlikely from a compositional perspective, Juno itself being classified as transitional L/LL \citep{Marsset:2024Nature}.

For context, the asteroid (25143) Itokawa, whose returned samples from the Hayabusa spacecraft indicate a thermally metamorphosed LL chondrite composition \citep{Nakamura:2011, Tsuchiyama:2011}, has a very high probability of a Flora origin (93.1\% in the two-family scenario and 88.4\% in the four-family one),
according to the NEOMOD model for kilometre-sized objects
\citep{Broz:2024,Broz:2026}. We note that ground-based spectroscopic observations correctly predicted its LL-chondrite composition well before the launch of Hayabusa \citep{Binzel:2001}.

\section{Thermal evolution and crust preservation}
\label{sec:thermal}

Differences in the petrologic-type distributions and their relative abundances between the Nysa\texorpdfstring{$_{\rm S}$}{S} and Flora families may reflect either distinct original parent-body sizes or different accretion times. 
To investigate this, we modelled the thermal evolution of LL-chondrite parent bodies with sizes comparable to those of the Nysa\texorpdfstring{$_{\rm S}$}{S} and Flora families using the spherically symmetric heat-conduction model of \citet{Monnereau:2013}. We then discuss the implications for the internal structure of the parent bodies and their present-day delivery of petrologic type-3 material to Earth.

\begin{figure*}[h!]
\centering
\includegraphics[width=\textwidth]{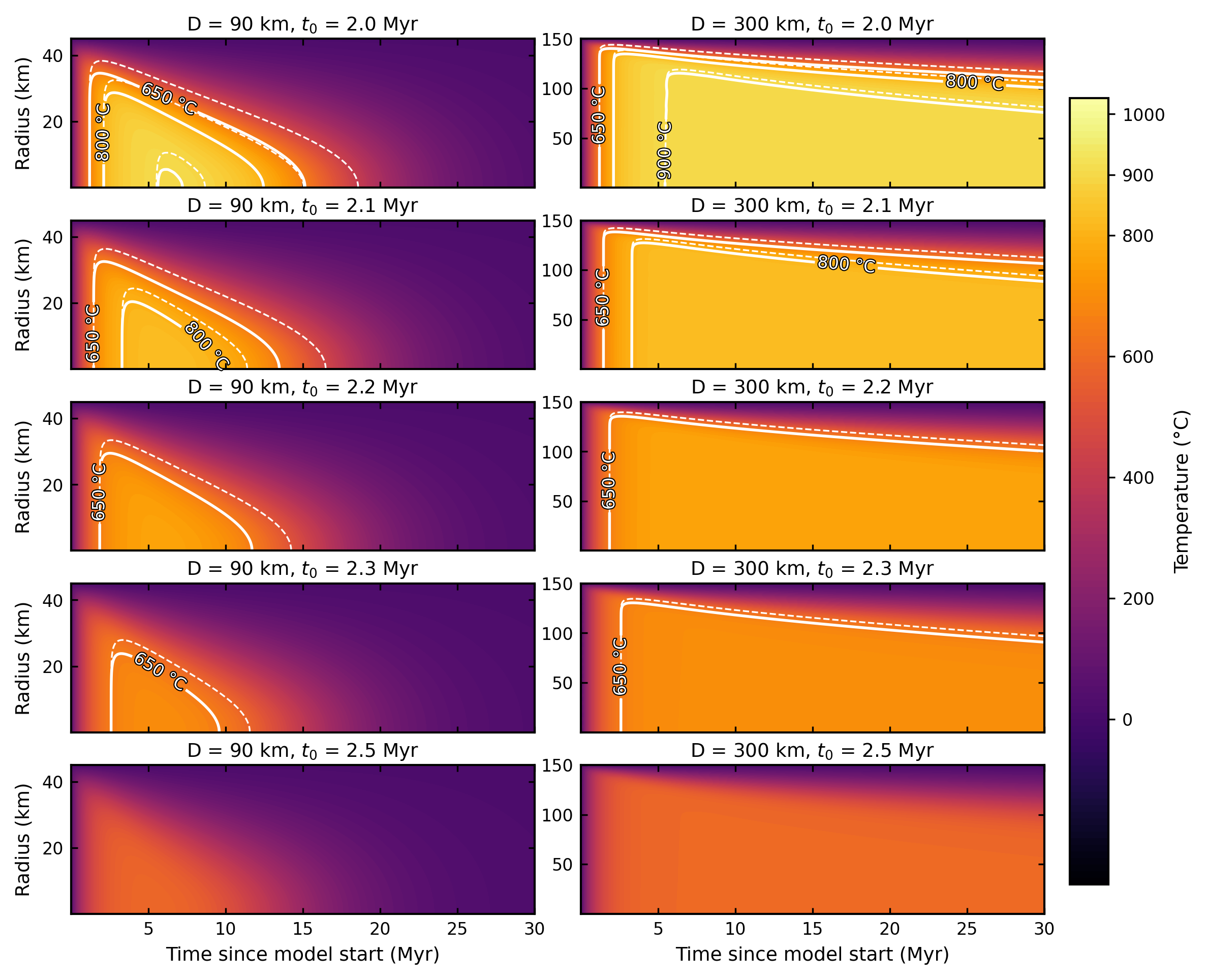}
\caption{{\bf Thermal evolution of LL-chondrite parent bodies with different sizes and accretion times.}
Temperature is shown as a function of time since model start and radial distance from the center, with colours indicating temperature (0--1300~K; see colour bar).
White contours mark the 650, 800, and 900$^\circ$C isotherms, corresponding to the boundaries between OC petrologic types~3/4, 4/5, and 5/6, respectively \citep{Edwards:2020}.
Thick solid and thin dashed contours indicate models with and without an insulating regolith of 150\,m thickness, respectively.
Panels show models for bodies with diameters of 90~km and 300~km.
We show accretion times of $t_0 = 2.0$, 2.1, 2.2, 2.3, and 2.5~Myr after CAI condensation.}
\label{fig:thermal}
\end{figure*}

\subsection{Sizes of LL-chondrite parent bodies}
\label{sec:sizes}

Several estimates of parent-body sizes, derived from collisional simulations and combined with subsequent mass loss inferred from N-body modelling, are available in the literature for the Nysa\texorpdfstring{$_{\rm S}$}{S} and Flora families \citep{Tanga:1999, Durda:2007, Broz:2013}. 
However, these estimates based on Smooth-Particle Hydrodynamics (SPH) simulations tend to underestimate parent-body sizes because they are calibrated to reproduce the observed family size–frequency distribution (SFD) only above the numerical resolution limit, while unresolved small fragments are implicitly excluded from the mass budget.
As a result, the inferred parent-body sizes are systematically underestimated.

A more robust estimate of parent-body sizes can be obtained by extrapolating the SFD of each family with a power law, $N = C D^{q}$, and integrating the corresponding volume distribution. 
Direct evidence that the SFDs of newly formed families extend down to sizes of order 100\,$\mu$m comes from meteorite statistics and zodiacal dust bands associated to the young Massalia$_2$, Karin, Koronis$_2$ and Veritas families \citep{Broz:2024, Broz:2024Nature, Marsset:2024Nature}.

Here, we consider either a single power law ($q_1$) or a broken power law ($q_1$, $q_2$) with a transition at diameter $D_0$.
Broken power laws are introduced to (1) avoid divergence of the total mass when extrapolating the SFD to small sizes, and (2) ensure that the extrapolated SFD remains consistent with that of the background main belt, as steeper distributions would be rapidly eroded by collisional cascades.

Observed SFDs are based on family identifications from \citet{Broz:2024Nature} (see their Supplementary Material), which use recent catalogues of orbital elements, albedos, and colours (AstOrb: \citealt{Moskovitz:2019}; AFP: \citealt{Knevzevic:2003, Novakovic:2019}; WISE: \citealt{Nugent:2016}; AKARI: \citealt{Usui:2011}; SDSS: \citealt{Parker:2008}). Families were identified using the hierarchical clustering method (HCM; \citealt{Zappala:1990}) applied to proper elements with a variable cut-off velocity, followed by optional halo inclusion and removal of interlopers. Interlopers were rejected based on physical properties ($p_V \in [0.1, 0.5]$, $a^\ast \in [-0.1, 0.5]$) and $H < H(a_p)$ \citep{Vokrouhlicky:2006a}, where
\begin{equation}
H(a_p) = 5 \log_{10} |a_p - a_0| / C.
\end{equation}
For the Nysa\texorpdfstring{$_{\rm S}$}{S} family, (135) Hertha (M-type) and its associated family are efficiently removed
with $a_0 = 2.42851\,\mathrm{au}$ and $C = 0.12 \times 10^{-4}\,\mathrm{au}$. The resulting sample consists predominantly of multi-kilometre bodies (up to $\sim$7 km).

The results of this procedure are presented in Table\,\ref{tab:pb_sizes}. In the case of the Flora family, the observed SFD for kilometre-sized members follows a power law with an index close to $-3.0$ \citep{Broz:2024Nature}. Extrapolating this relation down to 1\,km yields a parent-body diameter of $\sim$158\,km, while extending it further to 100\,$\mu$m increases the inferred size to $(196\pm 30)$\,km.

Similarly, we estimate the original size of the Nysa\texorpdfstring{$_{\rm S}$}{S} parent body to be $(90\pm 23)$\,km. Revised size estimates for the Koronis and Massalia parent bodies, $\sim$253 and $\sim$171\,km, respectively, are in better agreement with expectations from thermal models  (\citealt{Henke_2012A&A...545A.135H,Monnereau:2013,Blackburn:2017,Gail:2019}; see Table\,\ref{tab:pb_sizes}) compared to those inferred from SPH simulations.

\def\s#1{\textcolor{gray}{#1}}

\begin{figure*}
\centering
\includegraphics[width=\textwidth]{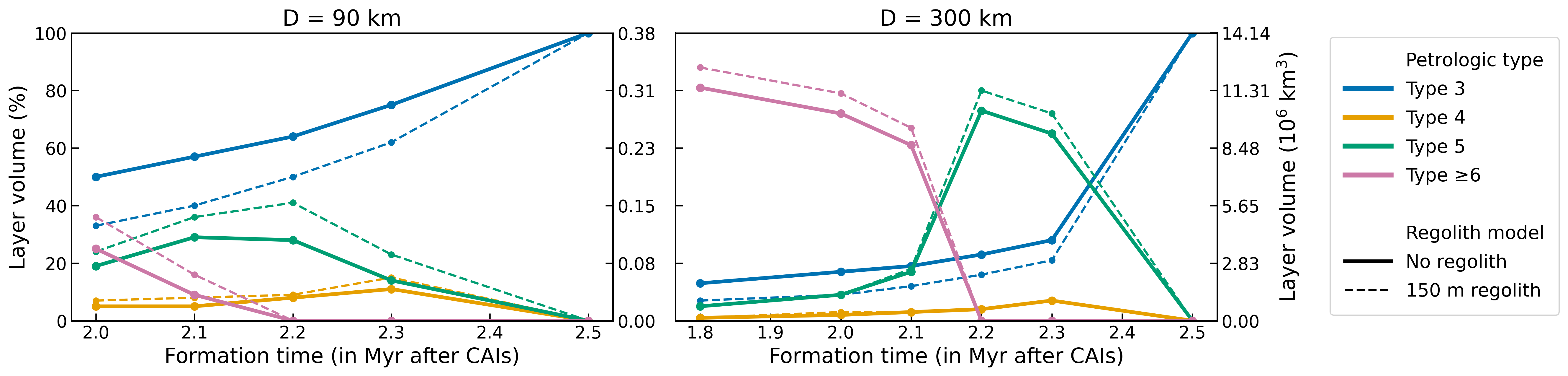}
\caption{{\bf Thermal model predictions for the internal structure of LL-chondrite parent bodies.}
For each parent-body diameter $D$, and for models with and without a 150 m thick insulating regolith, we report the volume of each petrologic layer as a function of formation time after CAI condensation. Volumes are given both in percent of the total body volume and in absolute units ($10^{6}$\,km$^{3}$).}
\label{fig:thermal2}
\end{figure*}

\subsection{Thermal history of the Nysa\texorpdfstring{$_{\rm S}$}{S} and Flora parent bodies}
\label{sec:pbs}

In the model of \citet{Monnereau:2013}, internal heating is dominated by the radioactive decay of $^{26}$Al and the thermal evolution is primarily controlled by two parameters: the time of accretion relative to the formation of Calcium–Aluminium-rich inclusions (CAIs; $t_0$), which sets the initial $^{26}$Al abundance, and a cooling timescale that depends on the body size and thermal properties. 
The model also accounts for temperature-dependent thermal conductivity and the presence of a low-conductivity surface regolith, which acts as an insulating layer and shifts isotherms toward the surface.

We adopt similar thermal parameters as \citet{Monnereau:2013} for the H-chondrite parent body, with adjustments to account for the physical properties of LL chondrites. Specifically, we assume a mean density of 3.21~g~cm$^{-3}$ (compared to 3.8~g~cm$^{-3}$ for H chondrites) and a metal mass fraction of $X_{\rm Fe}=0.10$ (compared to 0.23 for H chondrites).
In addition, following \citet{Edwards:2020}, we use an initial aluminium abundance of 1.18~wt\% and assume instantaneous accretion, consistent with recent dynamical models of both turbulent and coagulative accretion \citep{Johansen:2007, Cuzzi:2010, Weidenschilling:2011}.

Simulations are performed for parent bodies with diameters of 90 and 300~km, corresponding to our estimated size for the Nysa\texorpdfstring{$_{\rm S}$}{S} parent body, and the expected size of the parent body of equilibrated LL chondrites inferred from Pb-phosphate thermochronology \citep{Edwards:2020}, respectively. The latter also corresponds to the upper bound of our size estimate for the Flora parent body (Table~\ref{tab:pb_sizes}).
We explore a range of accretion times, from $t_0 = 1.8$ to 2.5~Myr after CAI condensation, for bodies both with and without a 150\,m-thick insulating regolith.

Figure~\ref{fig:thermal} illustrates the thermal evolution of these parent bodies as a function of radial distance and time since formation, and the corresponding stratigraphic structures are shown in Figure~\ref{fig:thermal2} and Appendix~\ref{sec:app_thermal}.

For formation times $t_0 \gtrsim 2.2$~Myr, even large parent bodies ($D \sim 300$~km) produce only type~3--4 material. The inclusion of an insulating regolith shifts isotherms toward the surface, but does not generate higher petrologic types. This implies that the parent bodies of EOCs -- including Flora and Eunomia -- must have accreted before $\sim$2.2~Myr.

Conversely, the Nysa\texorpdfstring{$_{\rm S}$}{S} parent body may have formed anytime after $\sim$2.0~Myr to account for the dominance of unequilibrated material within that family. Constraints from LL3 chondrule ages, which peak between $\sim$1.8 and 2.2~Myr after CAI condensation \citep{Hutcheon:1989,Kita:2000,Rudraswami:2007,Rudraswami:2008,Villeneuve:2009,Mishra:2010,Bollard:2019,Pape:2019,Schonbachler:2025}, and possibly within a narrower interval \citep{Siron:2022}, further indicate an accretion time of $\sim$2.1--2.2~Myr, as the parent body must incorporate this full range of formation ages.

Together, these constraints indicate that the Nysa\texorpdfstring{$_{\rm S}$}{S}, Flora, and Eunomia parent bodies could have formed contemporaneously, around $\sim$2.1--2.2~Myr after CAI condensation. Differences in parent-body size are sufficient to account for the observed petrologic variations, without invoking differences in formation time.

\subsection{Survival of type-3 material}
\label{sec:type3}

Thermal modelling further predicts that even the largest and most thermally metamorphosed OC parent bodies preserve an outer unequilibrated shell with a thickness of a few kilometres (Figure~\ref{fig:thermal2}).
Therefore, the absolute volume of petrologic type-3 material increases strongly with parent-body size.
For the parent bodies of Nysa\texorpdfstring{$_{\rm S}$}{S} ($D \approx 90$\,km) and Flora ($D \approx 300$\,km), the corresponding volumes of crustal (type-3) material are $\sim\,0.2\times10^{6}$\,km$^{3}$ and $(1.7$--$3.2)\times10^{6}$\,km$^{3}$, respectively (Appendix~\ref{sec:app_thermal}), depending on whether an insulating regolith is included.
Based on volume alone, one would therefore expect Flora and Eunomia to be the dominant present-day sources of type-3 material rather than Nysa\texorpdfstring{$_{\rm S}$}{S}.

However, the relevant quantity for meteorite delivery is not the total amount of type-3 material produced, but the fraction that can survive to the present day in deliverable fragments.
Collisional erosion during the planet-formation phase likely removed a significant fraction of the original unequilibrated crust prior to family-forming events \citep{Vernazza:2014}.
At the time of disruption, the maximum size of fragments composed entirely of type-3 material was therefore limited by the remaining crustal thickness, i.e. $\lesssim$12\,km (Appendix~\ref{sec:app_thermal}).
The Flora and Eunomia families are old, with estimated ages of $(1.2\pm0.2)$\,Gyr and $(4.2\pm0.3)$\,Gyr, respectively \citep[][SI]{Broz:2024Nature}, and collisional evolution models \citep{Bottke:2005_foss, Morbidelli:2009, Broz:2024Nature} show that fragments of this size cannot survive the collisional cascade over billion-year timescales. Consequently, most type-3 material produced during the formation of these old families is expected to have been collisionally destroyed.

On the other hand, the preservation of type-3 material in the Nysa\texorpdfstring{$_{\rm S}$}{S} family was likely enabled by two factors:
(1) reduced radiogenic heating due to the smaller size of the parent body, resulting in a thicker primordial type-3 shell; and, more importantly,
(2) the relatively young family age of $(0.6\pm0.1)$\,Gyr \citep{Broz:2024Nature}, for which multi-km-sized fragments have not yet been collisionally ground down.
Together, these factors allow crust-derived fragments to survive the collisional cascade and efficiently supply type-3 material to the present-day meteorite record.

\section{Conclusion}

The spectroscopic correspondence between unequilibrated (type~3) LL chondrites and the Nysa\texorpdfstring{$_{\rm S}$}{S} family, and between the most equilibrated (type~6-7) LL chondrites and the Flora family -- two families identified by dynamical models as efficient sources of NEOs and meteorites \citep{Broz:2024Nature} -- supports genetic links between these populations. 
This interpretation is further strengthened by the joint spectroscopic–dynamical correlation observed between LL-chondrite-like NEOs and the Nysa\texorpdfstring{$_{\rm S}$}{S} and Flora families, as well as, to a lesser extent, by the pre-atmospheric orbits of LL chondrites.

\begin{figure}
\centering
\includegraphics[width=\columnwidth]{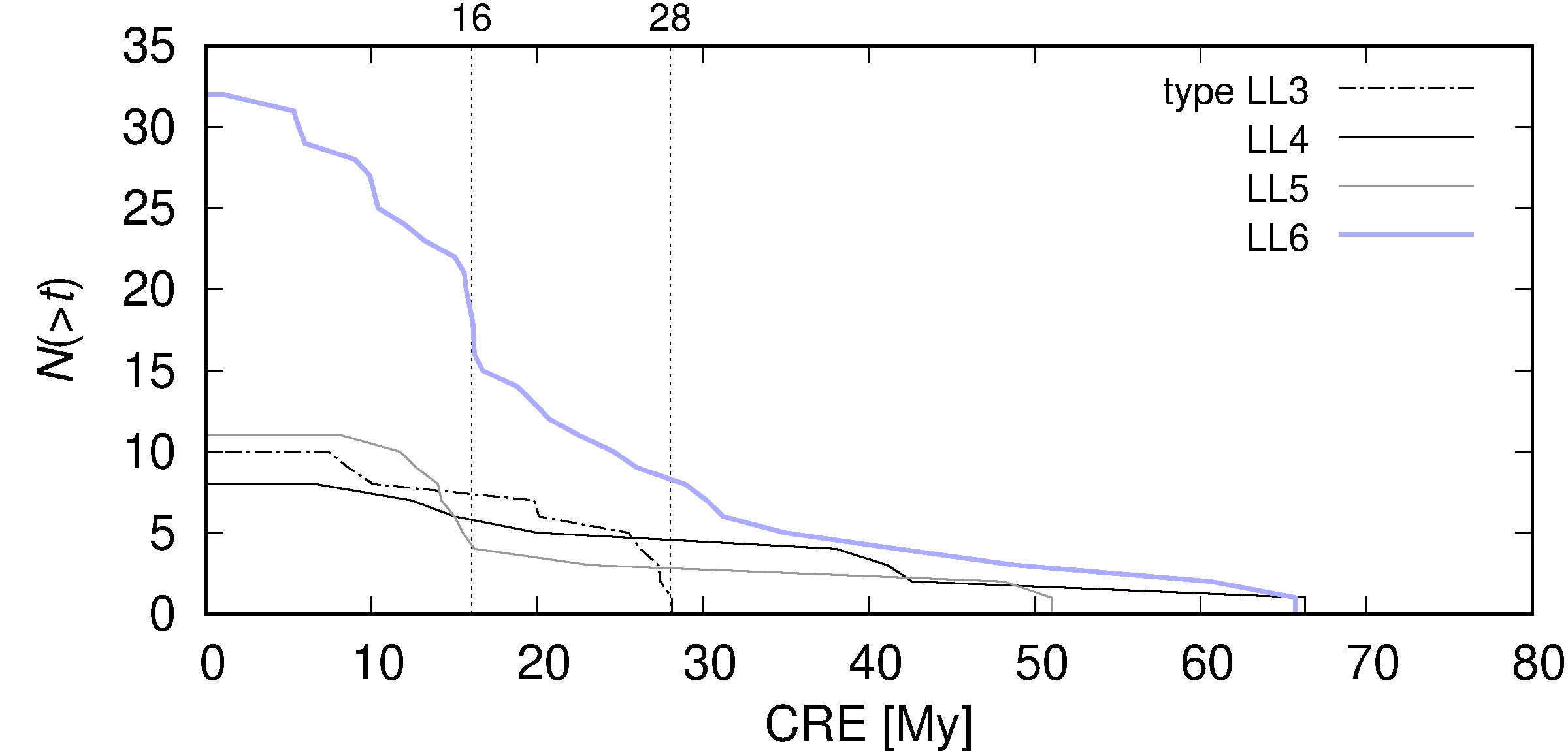}
\caption{{\bf Cosmic-ray exposure ages of LL chondrites.} 
The cumulative distributions are shown per petrologic type.}
\label{fig:cre}
\end{figure}

Different origins for low- and high-petrologic types of LL chondrites
are also consistent with their cosmic-ray exposure (CRE) ages
(Figure~\ref{fig:cre}; \citealt{Graf_1994Metic..29..643G}). 
In particular, a shared origin for most LL5 and LL6 chondrites
is strongly supported by their similar CRE age distribution,
with an onset at ${\sim}16$\,Myr,
indicative of a recent fragmentation event,
occurring within an ongoing collisional cascade in Flora.
LL4 chondrites display a relatively flat CRE-age distribution, extending up to ${\sim}66$~Myr, which may indicate an older fragmentation event within either the Nysa\texorpdfstring{$_{\rm S}$}{S} or Flora families.
In contrast, LL3 chondrites exhibit a different distribution, with ages $\lesssim28$\,Myr \citep{Eugster:2006},
consistent with a distinct, Nysa\texorpdfstring{$_{\rm S}$}{S} source.

Constraints from the melting ages of LL3 chondrules (\citealt{Schonbachler:2025} and references therein) indicate an accretion age of $\sim$2.1--2.2\,Myr after CAI condensation for the LL3 parent body. Thermal modelling further shows that, for formation times around $\sim$2.1\,Myr, large parent bodies ($D \sim 300$~km) still produce substantial fractions of equilibrated material, consistent with the observed mineralogy of the Flora and Eunomia families. Petrologic differences between the Nysa\texorpdfstring{$_{\rm S}$}{S}, Flora, and Eunomia parent bodies can therefore be naturally explained by differences in parent-body size, without requiring distinct formation times, although a delay of $\lesssim$0.2\,Myr cannot be excluded.


\section*{Code and data availability}

The IDL implementation of the Shkuratov radiative-transfer model used in this work is publicly available at \url{https://github.com/mmarsset/Shkuratov-IDL/}.

The spectra and full electronic versions of the appendix tables presented in this work are publicly available on Zenodo at \url{https://doi.org/10.5281/zenodo.20080259}.

\begin{acknowledgements}

The authors warmly thank Graham H. Edwards and Sunao Hasegawa for their careful reviews of this work, and Jolantha Eschrig for valuable discussions on the classification of ordinary chondrites.
M.B. and J.H. were supported by GACR grant no. 25-16789S of the Czech Science Foundation. F.E.D. acknowledges NASA grant 80NSSC22K0773.
Observations reported here were obtained at the
NASA Infrared Telescope Facility, which is operated by the University of Hawaii under contract 80HQTR19D0030 with the National Aeronautics and Space Administration. The authors acknowledge the sacred nature of Maunakea and appreciate the opportunity to observe from the mountain.

\end{acknowledgements}

\bibliographystyle{bibtex/aa}
\bibliography{references}

\begin{appendix}

\onecolumn

\section{Nomenclature of the Nysa complex}
\label{sec:app_names}

The taxonomically diverse and dynamically overlapping families of the Nysa-Polana complex have undergone a complex history of identification and redefinition, with evolving membership, taxonomy, and nomenclature \citep{Dykhuis:2015}. While Polana (and Eulalia) are now clearly recognised as distinct B- and C-type families \citep{Walsh:2013}, the S-type component and the embedded X-type cluster -- referred to in this work as Nysa\texorpdfstring{$_{\rm S}$}{S} and Hertha, respectively -- have been identified under different names in the literature. Table~\ref{tab:family_names} summarises these various designations to connect these works.

\begin{table*}[h!]
\centering
\caption{Summary of the various names of the Nysa\texorpdfstring{$_{\rm S}$}{S} and Hertha families used in the literature.}
\label{tab:family_names}
\begin{tabular}{lccccccccc}
\hline
This work & Tax. & Z77 & W79 & B89 & Z95 & C01 & M05 & M14 &... \\
\hline
Nysa\texorpdfstring{$_{\rm S}$}{S}  & S & Hertha & W-160 & Hertha & Nysa & Mildred & Mildred/Hertha & Burdett & \\
Hertha  & X & -- & -- & -- & -- & -- & Hertha & -- & \\
\hline
\end{tabular}
\\
\vspace{2mm}

\begin{tabular}{cccccccc}
\hline
... & D15 & M15 & N15 & B24 & M24 & N24 & J25 \\
\hline
 & Hertha1 & Nysa/Hertha & Nysa (405) & Nysa & Nysa & Nysa & Hertha \\
 & Hertha2 & --          & --   & Hertha & -- & Hertha & Hertha${_2}$ \\
\hline
\end{tabular}

\tablefoot{References: Z77=\citet{Zellner:1977}; 
W79=\citet{Williams:1979};
B89=\citet{Bell:1989};
Z95=\citet{Zappala:1995};
C01=\citet{Cellino:2001};
M05=\citet{Mothe:2005};
W13=\citet{Walsh:2013};
M14=\citet{Milani:2014};
D15=\citet{Dykhuis:2015};
M15=\citet{Masiero:2015};
N15=\citet{Nesvorny:2015_families};
B24=\citet{Broz:2024Nature};
M24=\citet{Marsset:2024Nature};
N24=\citet{Nesvorny:2024};
J25=\citet{Jenniskens:2025}.}

\end{table*}

\section{Meteorite dataset}
\label{sec:app_met}

Table~\ref{tab:met} lists the 133 laboratory spectra of 70 unique LL chondrites from the RELAB (99 spectra) and SSHADE (34 spectra) databases considered in this work.
For each spectrum, we provide the meteorite name, petrologic type, database of origin, and spectral file name.

For the SSHADE dataset, we adopted the petrologic classifications of \citet{Bonal:2016} based on the structural order of the polyaromatic carbonaceous matter assessed by Raman spectroscopy, and the revised H, L, and LL compositional classification of \citet{Eschrig:2022}. 
This classification is based on metal abundance, determined through point counting and magnetic susceptibility measurements, which directly probe ferromagnetic phases and show minimal overlap between OC groups while being largely insensitive to metamorphic grade \citep{Rochette:2003}. In contrast, classifications in the MBD often rely on Mg\# measurements of olivine and pyroxene, which are more difficult to apply to UOCs due to their small grain sizes and heterogeneous compositions. For these reasons, we consider the classification of \citet{Eschrig:2022} to be more reliable and adopt it in this work.

The petrologic class distribution derived from the MBD (Figure\,\ref{fig:histo1}) was corrected by accounting for the fractional reclassification of meteorites into and out of each class and applying these fractions to the MBD distribution.
Let $X_{H\rightarrow LL}$, $X_{L\rightarrow LL}$, $X_{LL\rightarrow L}$, and $X_{LL\rightarrow H}$ denote the fractions of type-3 OCs reclassified from H to LL, from L to LL, from LL to L, and from LL to H, respectively, as determined by \citet{Eschrig:2022}.
The corrected number of LL3 chondrites is therefore estimated as:
\begin{equation}
N_{\rm LL3}^{\rm corr} = \left(1 + X_{H\rightarrow LL} + X_{L\rightarrow LL} - X_{LL\rightarrow L} - X_{LL\rightarrow H}\right)\,N_{\rm LL3}.
\end{equation}
In \citet{Eschrig:2022}'s sample of 41 type-3 OCs, these fractions correspond to $X_{H\rightarrow LL}=0.098$, $X_{L\rightarrow LL}=0.268$, $X_{LL\rightarrow L}=0.073$, and $X_{LL\rightarrow H}=0.000$.
The same correction procedure is applied to the H3 and L3 populations.
Note that, for simplicity and to remain conservative (i.e., to avoid overestimating reclassifications from H and L to LL), we assign L(LL) meteorites to the L category and LL(L) meteorites to the LL category, while the intermediate L/LL Bishunpur meteorite is also considered as L.


\begin{table*}[ht]
\centering
\tiny
\setlength{\tabcolsep}{3pt}
\renewcommand{\arraystretch}{1.05}

\caption{LL chondrites considered in this work.}
\label{tab:met}

\begin{tabular}{l l l l r r l c}
\hline\hline
Meteorite & Type & Database & Spectral file &
$\chi^2_{\mathrm{Nysa}}$ &
$\chi^2_{\mathrm{Flora}}$ &
best & $Q$ \\
\hline

Aldsworth & LL5 & RELAB & hym/mt/c1mt77.txt & 4.04e-05 & 4.25e-04 & Nysa & g \\
Allan Hills 83007 & LL3.4 & RELAB & pfv/mt/c1mt136a.txt & 3.11e-04 & 1.01e-03 & Nysa & p \\
Allan Hills 83008 & LL3.4--3.7 & SSHADE & refl-NIR-ALH83008\_T80C\_i0\_e30.data.txt & 3.33e-04 & 1.16e-03 & Nysa & p \\
Allan Hills 83010 & LL3.3 & RELAB & pfv/mt/c1mt138a.txt & 1.65e-04 & 6.07e-04 & Nysa & g \\
Allan Hills 84086 & LL3 & SSHADE & refl-NIR-ALH84086\_T80C\_i0\_e30.data.txt & 1.51e-04 & 1.74e-04 & Nysa & g \\
Allan Hills 84096 & LL6 & RELAB & txh/mb/c1mb83.txt & 2.32e-04 & 6.72e-04 & Nysa & p \\
Allan Hills 84126 & LL3.4 & RELAB & pfv/mt/c1mt170a.txt & 2.84e-04 & 2.01e-04 & Flora & p \\
Allan Hills A77278 & LL3.7 & RELAB & pfv/mt/bir2mt156a.txt & 1.92e-04 & 4.09e-04 & Nysa & g \\
Allan Hills A77278 & LL3.7 & RELAB & pfv/mt/bir2mt156b.txt & 6.68e-04 & 2.01e-03 & Nysa & p \\
Allan Hills A77278 & LL3.7 & RELAB & pfv/mt/c1mt156a.txt & 1.63e-04 & 2.40e-04 & Nysa & g \\
Allan Hills A77278 & LL3.7 & RELAB & pfv/mt/c2mt156a.txt & 1.56e-04 & 2.63e-04 & Nysa & g \\
Allan Hills A77278 & LL3.7 & RELAB & pfv/mt/c2mt156b.txt & 4.72e-04 & 1.57e-03 & Nysa & p \\
Allan Hills A78119 & LL3.5 & RELAB & pfv/mt/c1mt147a.txt & 2.19e-04 & 7.67e-04 & Nysa & p \\
Alta'ameem & LL5 & RELAB & hym/mt/c1mt78.txt & 2.02e-04 & 3.19e-04 & Nysa & p \\
Alta'ameem & LL5 & RELAB & txh/oc/c1oc10a.txt & 2.38e-03 & 1.28e-03 & Flora & p \\
Alta'ameem & LL5 & RELAB & txh/oc/c1oc10b.txt & 1.83e-03 & 1.17e-03 & Flora & p \\
Alta'ameem & LL5 & RELAB & txh/oc/c1oc10c.txt & 2.13e-03 & 1.42e-03 & Flora & p \\
Alta'ameem & LL5 & RELAB & txh/oc/c2oc10b.txt & 1.45e-03 & 1.10e-03 & Flora & p \\
Alta'ameem & LL5 & RELAB & txh/oc/c2oc10c.txt & 2.97e-04 & 8.70e-04 & Nysa & p \\
Alta'ameem & LL5 & RELAB & txh/oc/c3oc10b.txt & 1.39e-03 & 1.02e-03 & Flora & p \\

\hline
\end{tabular}

\tablefoot{Full table available on Zenodo\footnote{\label{fn:zenodolink}\url{https://doi.org/10.5281/zenodo.20080259}}.}

\end{table*}

\section{Main-belt dataset}
\label{sec:app_shku}

Table~\ref{tab:shku} presents the list of Flora (44 members), Eunomia (16 members), and Nysa\texorpdfstring{$_{\rm S}$}{S} (13 members) asteroid family samples analysed in this work.
For each object, we report the asteroid number, name, and designation; the associated spectral file; the orbital elements ($a$, $e$, $i$); the absolute magnitude ($H$); and the results of compositional modelling based on their reflectance spectra (47, 16, and 14 spectra, respectively), obtained using the radiative transfer model of \citet{Shkuratov:1999}.
The fitted compositional parameters include the olivine and orthopyroxene abundances (\textit{ol} and \textit{opx}), the grain size (\textit{GS}, in $\mu$m), and the space-weathering coefficient ($c_s$) \citep{Brunetto:2006}.

\begin{figure}[h]
\centering
\includegraphics[clip, width=0.5\columnwidth]{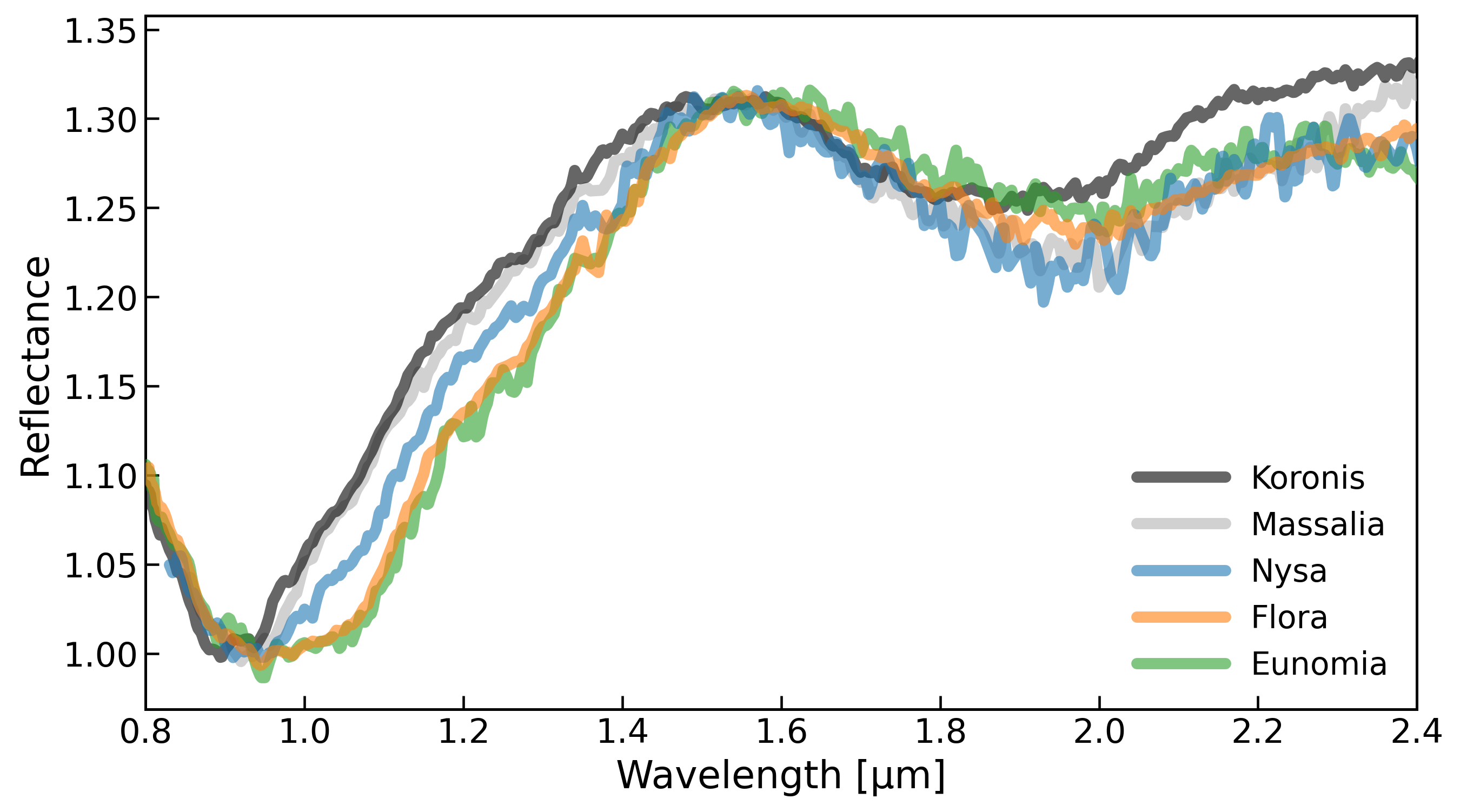}
\caption{{\bf Spectral comparison of ordinary chondrite parent bodies.} 
We present the average spectra of asteroid families associated with H, L, unequilibrated LL, and equilibrated LL chondrites: Koronis \citep{Broz:2024Nature}, Massalia \citep{Marsset:2024Nature}, Nysa (this work), and Flora \citep{Vernazza:2008, Broz:2024Nature}, respectively. We also include Eunomia, which is discussed in this work. All spectra were dereddened following the procedure described in Section\,\ref{sec:spectra} and normalised to unity at the minimum of the $\sim$1~$\mu$m absorption band.}
\label{fig:mba_sptpb}
\end{figure}

Figure~\ref{fig:mba_sptpb} shows the average spectra of asteroid families associated with OCs, based on this work and previous results \citep{Vernazza:2008, Broz:2024Nature, Marsset:2024Nature}: Koronis (H chondrites), Massalia (L), Juno (L/LL), Nysa\texorpdfstring{$_{\rm S}$}{S} (LL3), Flora (equilibrated LL), and Eunomia (equilibrated LL). 
Average family spectra were obtained by computing a weighted mean of the member spectra, using the squared measurement uncertainties as weights, after removing the overall spectral slope using the empirical exponential function of \citet{Brunetto:2006} to account for space-weathering variations across family members. 
These families exhibit a progressive increase in the olivine-to-pyroxene ratio, expressed as a systematic shift of the $\sim$1~$\mu$m absorption band toward longer wavelengths (from $\sim$0.9 to $\sim$1.0~$\mu$m), along with the emergence of a secondary absorption feature near 1.3~$\mu$m attributed to olivine. A similar trend is observed among the corresponding meteorite groups, from H to L to LL.

\begin{figure}[h]
\centering
\includegraphics[clip, width=0.65\columnwidth]{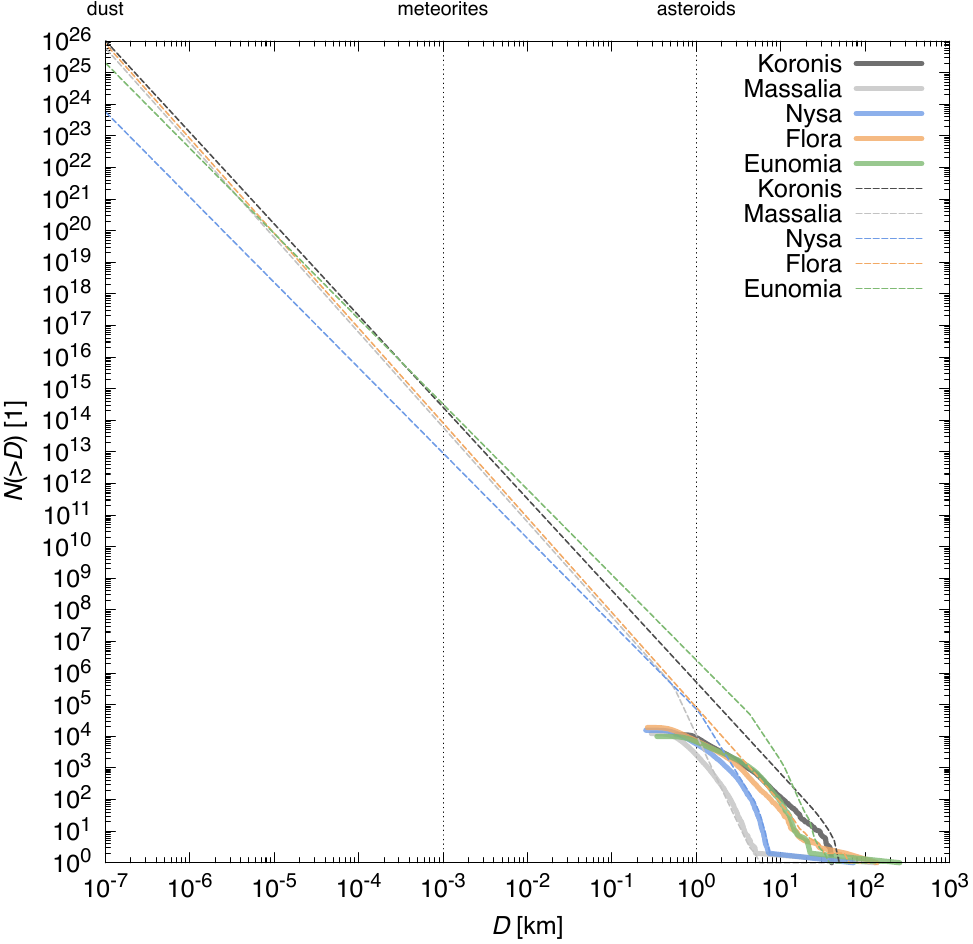}
\caption{
{\bf Size–frequency distributions (SFDs) used to derive the parent-body sizes of asteroid families.}
For each asteroid family, the currently observed SFD, derived as described in Section\,\ref{sec:sizes}, is shown as a thick line, while the preferred extrapolation -- representing the SFD at the time of family formation and based on the power-law slopes listed in Table\,\ref{tab:pb_sizes} -- is shown as a thin dashed line. The parent-body sizes from these extrapolations are in better agreement with expectations from thermal models of OC parent bodies than those obtained from traditional SPH simulations \citep{Durda:2007}.
}
\label{fig:mba_sfds}
\end{figure}

The SFDs shown in Figure~\ref{fig:mba_sfds} illustrate both the currently observed populations and their extrapolation toward smaller sizes, representing the full size distribution at the time of family formation. The extrapolated distributions are constrained by the slopes derived from the observed kilometre-sized population and by independent evidence that young asteroid families extend down to sub-millimetre sizes \citep{Broz:2024, Broz:2024Nature, Marsset:2024Nature}.

Despite both being associated with LL chondrites, the distinct petrologic types inferred for the Nysa\texorpdfstring{$_{\rm S}$}{S} and Flora families argue against a common progenitor (“grandparent body”). Such a scenario would require highly selective disruption (sampling crust versus interior) as well as a parent-body size inconsistent with thermal evolution models (e.g., \citealt{Edwards:2020}). It is also dynamically implausible: the Nysa\texorpdfstring{$_{\rm S}$}{S} parent body cannot be a fragment of the older Flora family, as the two families are well separated in orbital element space ($\sim 2.20$ vs.\ $\sim 2.42$~au), and an object of this size could not have migrated across such distances via Yarkovsky drift \citep{Vokrouhlicky:2006a} or chaotic diffusion over the age of the Solar System. Conversely, an early disruption followed by implantation into distinct regions of the main belt during the epoch of planet formation and migration \citep{Raymond:2014} would require a contrived scenario for which there is currently no supporting evidence.


\begin{table*}[ht]
\centering
\tiny
\setlength{\tabcolsep}{3pt}
\renewcommand{\arraystretch}{1.05}

\caption{Members of the Flora, Eunomia, and Nysa$_{\rm S}$ families considered in this work.}
\label{tab:shku}

\begin{tabular}{r l l l r r r r l r r r r r}
\hline\hline
Number & Name & Designation & Spectral file &
$a$ & $e$ & $i$ & $H$ & Family &
$\mathrm{ol}/(\mathrm{ol}+\mathrm{opx})$ &
ol & opx & GS & $c_s$ \\
\hline

8 & Flora & A847 UA & a000008.sp17.txt & 2.201 & 0.157 & 5.89 & 6.61 & Flora & 0.8294 & 3.4022 & 0.6996 & 7.9193 & 0.1976 \\
43 & Ariadne & A857 GA & a000043.sp48.txt & 2.203 & 0.169 & 3.47 & 7.94 & Flora & 0.8011 & 2.7635 & 0.6861 & 10.1975 & 0.1630 \\
254 & Augusta & A886 FA & a254.txt & 2.195 & 0.122 & 4.51 & 11.85 & Flora & 0.9317 & 11.9183 & 0.8741 & 1.0001 & 0.7107 \\
281 & Lucretia & A888 UC & a281.txt & 2.188 & 0.133 & 5.30 & 12.05 & Flora & 0.8115 & 8.6399 & 2.0070 & 9.0385 & 0.4580 \\
352 & Gisela & A893 AB & a000352.sp06.txt & 2.194 & 0.149 & 3.38 & 10.10 & Flora & 0.8378 & 3.5644 & 0.6901 & 12.1003 & 0.2022 \\
364 & Isara & A893 FE & a000364.sp61.txt & 2.221 & 0.150 & 6.00 & 9.92 & Flora & 0.8343 & 4.0809 & 0.8104 & 12.7183 & 0.3359 \\
364 & Isara & A893 FE & a000364.sp64.txt & 2.221 & 0.150 & 6.00 & 9.92 & Flora & 0.8311 & 5.7696 & 1.1726 & 7.9245 & 0.2286 \\
453 & Tea & A900 DD & a000453.sp07.txt & 2.183 & 0.109 & 5.55 & 10.57 & Flora & 0.8518 & 4.3637 & 0.7594 & 9.3861 & 0.1796 \\
913 & Otila & A919 KD & a000913.sp07.txt & 2.197 & 0.170 & 5.81 & 12.03 & Flora & 0.8138 & 3.3132 & 0.7580 & 11.8565 & 0.2761 \\
929 & Algunde & A920 ED & a000929.sp06.txt & 2.239 & 0.114 & 3.91 & 11.67 & Flora & 0.7969 & 4.2973 & 1.0951 & 7.0686 & 0.1720 \\
1185 & Nikko & 1927 WC & a1185.txt & 2.237 & 0.106 & 5.70 & 12.17 & Flora & 0.8004 & 6.0258 & 1.5024 & 6.4343 & 0.5182 \\
1412 & Lagrula & 1937 BA & a001412.dm02.txt & 2.214 & 0.113 & 4.72 & 12.48 & Flora & 0.8223 & 4.4399 & 0.9595 & 10.5186 & 0.2487 \\
1667 & Pels & 1930 SY & a001667.sp19.txt & 2.190 & 0.156 & 4.62 & 11.97 & Flora & 0.8172 & 2.2460 & 0.5024 & 10.2849 & 0.3186 \\
1807 & 1807 Slovakia & 1971 QA & a001807.sp30.txt & 2.226 & 0.178 & 3.49 & 12.58 & Flora & 0.7490 & 3.6198 & 1.2130 & 5.2574 & 0.4554 \\
1857 & 1857 Parchomenko & 1971 QS1 & a001857.sp263n1.txt & 2.243 & 0.135 & 4.40 & 12.34 & Flora & 0.8376 & 4.1549 & 0.8054 & 14.3782 & 0.4093 \\
2088 & 2088 Sahlia & 1976 DJ & a2088.txt & 2.207 & 0.079 & 5.54 & 12.86 & Flora & 0.8056 & 3.1374 & 0.7571 & 7.2974 & 0.3396 \\
2171 & Kiev & 1973 QD1 & a2171.txt & 2.256 & 0.166 & 7.52 & 12.82 & Flora & 0.8470 & 3.0865 & 0.5574 & 8.5591 & 0.2372 \\
2647 & Sova & 1980 SP & a2647.txt & 2.244 & 0.138 & 3.94 & 12.84 & Flora & 0.8201 & 3.0795 & 0.6754 & 10.8725 & 0.3306 \\
2873 & Binzel & 1982 FR & a002873.sp09.txt & 2.251 & 0.158 & 5.90 & 13.06 & Flora & 0.8354 & 1.5978 & 0.3147 & 16.5301 & 0.0565 \\
3029 & Sanders & 1981 EA8 & a003029.sp97.txt & 2.240 & 0.112 & 3.42 & 13.00 & Flora & 0.8710 & 8.5068 & 1.2604 & 9.7422 & 0.3346 \\

\hline
\end{tabular}

\tablefoot{Full table available on Zenodo\hyperref[fn:zenodolink]{\textsuperscript{\ref*{fn:zenodolink}}}.}

\end{table*}

\section{NEO dataset}
\label{sec:app_neo}

Table~\ref{tab:app_neo} lists all 395 spectra of 325 individual LL-chondrite-like NEOs from the MITHNEOS survey \citep{Binzel:2019,Marsset:2022} considered in this study. 
For each object, we report the asteroid number, name, and designation; the associated spectral file; the orbital elements ($a$, $e$, $i$); the absolute magnitude ($H$); and the measured ol/(ol+opx) ratio.


\begin{table*}[ht]
\centering
\tiny
\setlength{\tabcolsep}{3pt}
\renewcommand{\arraystretch}{1.05}

\caption{LL-chondrite-like NEOs considered in this work.}
\label{tab:app_neo}

\begin{tabular}{llllrrrrrrrl}
\hline\hline
Number & Name & Designation & Spectral file &
$a$ [au] & $e$ & $i$ [$^\circ$] & $H$ &
$\mathrm{ol}/(\mathrm{ol}+\mathrm{opx})$ &
$\chi^2_{\nu}$ (Nysa$_{\rm S}$) &
$\chi^2_{\nu}$ (Flora) & best \\
\hline
433 & Eros & 1898 DQ & a000433.sp101.txt & 1.458 & 0.223 & 10.83 & 10.38 & 0.788 & 5.90e-04 & 4.31e-05 & Flora \\
433 & Eros & 1898 DQ & a000433.sp102.txt & 1.458 & 0.223 & 10.83 & 10.38 & 0.784 & 6.07e-04 & 5.40e-05 & Flora \\
433 & Eros & 1898 DQ & a000433.sp103.txt & 1.458 & 0.223 & 10.83 & 10.38 & 0.793 & 7.30e-04 & 1.18e-04 & Flora \\
433 & Eros & 1898 DQ & a000433.sp105.txt & 1.458 & 0.223 & 10.83 & 10.38 & 0.830 & 9.77e-04 & 2.11e-04 & Flora \\
433 & Eros & 1898 DQ & a000433.sp15.txt & 1.458 & 0.223 & 10.83 & 10.38 & 0.743 & 2.98e-04 & 7.55e-05 & Flora \\
433 & Eros & 1898 DQ & a000433.sp16.txt & 1.458 & 0.223 & 10.83 & 10.38 & 0.770 & 3.92e-04 & 2.81e-05 & Flora \\
433 & Eros & 1898 DQ & a000433.sp17.txt & 1.458 & 0.223 & 10.83 & 10.38 & 0.757 & 3.04e-04 & 6.15e-05 & Flora \\
433 & Eros & 1898 DQ & a000433.sp201n1.txt & 1.458 & 0.223 & 10.83 & 10.38 & 0.765 & 2.86e-04 & 1.38e-04 & Flora \\
433 & Eros & 1898 DQ & a000433.sp223.txt & 1.458 & 0.223 & 10.83 & 10.38 & 0.765 & 3.49e-04 & 2.81e-05 & Flora \\
433 & Eros & 1898 DQ & a000433.sp245n2.txt & 1.458 & 0.223 & 10.83 & 10.38 & 0.778 & 5.19e-04 & 8.35e-05 & Flora \\
433 & Eros & 1898 DQ & a000433.sp247n1.txt & 1.458 & 0.223 & 10.83 & 10.38 & 0.785 & 3.60e-04 & 9.30e-05 & Flora \\
433 & Eros & 1898 DQ & a000433.sp256n2.txt & 1.458 & 0.223 & 10.83 & 10.38 & 0.788 & 4.42e-04 & 1.02e-04 & Flora \\
1566 & Icarus & 1949 MA & a001566.sp42.txt & 1.078 & 0.827 & 22.80 & 16.53 & 0.806 & 1.97e-03 & 7.64e-04 & Flora \\
1620 & Geographos & 1951 RA & a001620.sp03.txt & 1.246 & 0.336 & 13.34 & 15.26 & 0.792 & 2.95e-04 & 8.69e-05 & Flora \\
1620 & Geographos & 1951 RA & a001620.sp105.txt & 1.246 & 0.336 & 13.34 & 15.26 & 0.814 & 5.36e-04 & 5.44e-05 & Flora \\
1620 & Geographos & 1951 RA & a001620.sp68.txt & 1.246 & 0.336 & 13.34 & 15.26 & 0.834 & 9.14e-04 & 1.67e-04 & Flora \\
1627 & Ivar & 1929 SH & a001627.sp09.txt & 1.863 & 0.397 & 8.46 & 12.79 & 0.789 & 1.90e-04 & 3.78e-04 & Nysa \\
1627 & Ivar & 1929 SH & a001627.sp118n1.txt & 1.863 & 0.397 & 8.46 & 12.79 & 0.795 & 9.44e-04 & 3.18e-04 & Flora \\
1627 & Ivar & 1929 SH & a001627.sp243n2.txt & 1.863 & 0.397 & 8.46 & 12.79 & 0.758 & 2.64e-04 & 1.55e-04 & Flora \\
1627 & Ivar & 1929 SH & a001627.sp246n1.txt & 1.863 & 0.397 & 8.46 & 12.79 & 0.775 & 3.19e-04 & 2.29e-04 & Flora \\
\hline
\end{tabular}

\tablefoot{Full table available on Zenodo\hyperref[fn:zenodolink]{\textsuperscript{\ref*{fn:zenodolink}}}.}

\end{table*}

\section{Spectral comparisons}
\label{sec:app_comp}

Tables~\ref{tab:met} and \ref{tab:app_neo} further report the $\chi^2$ values obtained from comparisons of meteorite and NEO spectra, respectively, with the mean Nysa\texorpdfstring{$_{\rm S}$}{S} and Flora family spectra, and used to identify their closest spectral analogues. All spectra were first normalised at the wavelength of the 1\,$\mu$m band minimum and dereddened using the LL-chondrite template of \citet{Vernazza:2008} as a reference, such that the reflectance maximum near 1.5-1.6\,$\mu$m matches that of the template.

To reduce sensitivity to noise and small-scale spectral features, the wavelengths of the reflectance minimum and maximum were determined by locally fitting a polynomial and extracting the corresponding extrema. For high-SNR meteorite spectra, a sixth-order polynomial was adopted. For NEO spectra, which span a wider range of SNRs, the polynomial order was allowed to vary between third and sixth order, and the fit minimising the $\chi^2$ was retained.

$\chi^2$ values were calculated over the 0.90–1.75\,$\mu$m range, which encompasses the 1\,$\mu$m band -- the most diagnostic feature of silicate composition, particularly the olivine-to-pyroxene ratio -- while retaining some sensitivity to the 2\,$\mu$m band depth. The upper limit was chosen to avoid excessive sensitivity to variations in meteorite sample preparation and space-weathering effects, which may not be fully captured by the simple exponential model used here.

The $\chi^2$ values are unweighted and not reduced, as measurement uncertainties are not available for RELAB spectra; they are used solely to identify the closest spectral analogue rather than as formal goodness-of-fit metrics. In the case of meteorites, a $\chi^2$ threshold was determined from visual inspection of the spectral fits to distinguish good from poor matches.

For the RELAB dataset, we adopt a threshold of $2 \times 10^{-4}$, and for SSHADE $3.2 \times 10^{-4}$. The difference between these thresholds reflects the different average SNRs of the two datasets, with RELAB spectra having higher SNR on average. 
The "Q" column in Table~\ref{tab:met} indicates whether a given spectral comparison meets the quality threshold (“g” for good, “p” for poor), while the best-matching family is reported in the “best” column. LL3, and LL6 or 7 chondrites classified as “g” are the ones used to compute the average spectra in Figure\,\ref{fig:meteorites}.

\begin{figure*}[h!]
\centering
\includegraphics[width=0.9\textwidth]{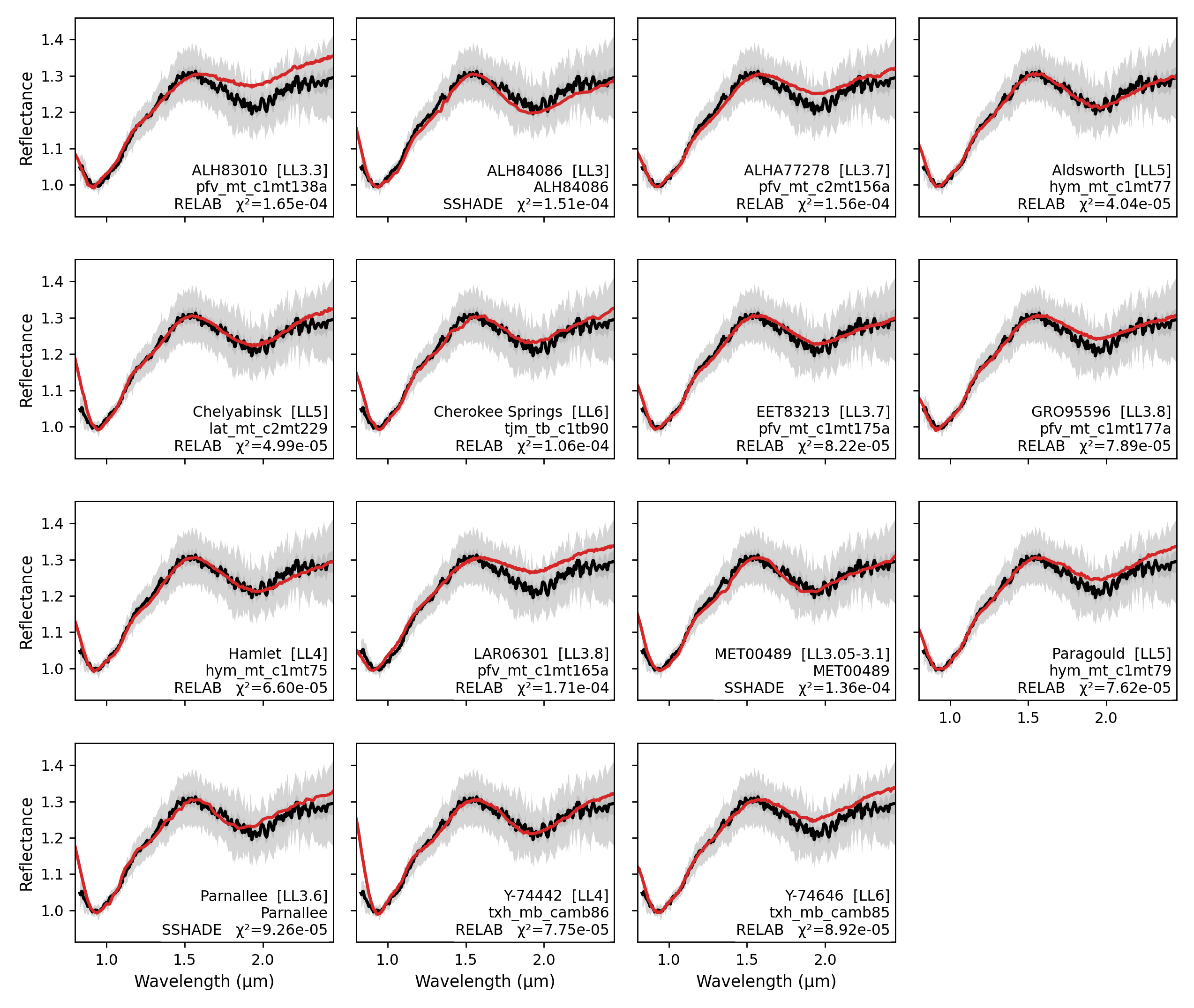}
\caption{{\bf Spectral comparison of LL chondrites with the Nysa\texorpdfstring{$_{\rm S}$}{S} family.} Shown is a selection of LL chondrite spectra (red) exhibiting good agreement with the mean Nysa\texorpdfstring{$_{\rm S}$}{S} family spectrum (black, with the grey envelope indicating the standard deviation among family members). For each spectrum, the meteorite name, petrologic type, file name (shortened in the case of SSHADE spectra), data source (RELAB or SSHADE), and the corresponding $\chi^2$ value (computed as described in Appendix\,\ref{sec:app_comp}) are indicated.
All spectra were dereddened using the LL-chondrite template of \citet{Vernazza:2008} as a reference. 
The most diagnostic feature is the 1\,$\mu$m absorption band. 
Divergence beyond $\sim$1.9\,$\mu$m can be attributed to meteorite sample preparation and differences in space-weathering levels, which are not fully captured by the simple exponential model used here.}
\label{fig:met_nysa}
\end{figure*}

\begin{figure*}[h!]
\centering
\includegraphics[trim={0 0 0 0}, clip, width=0.9\textwidth]{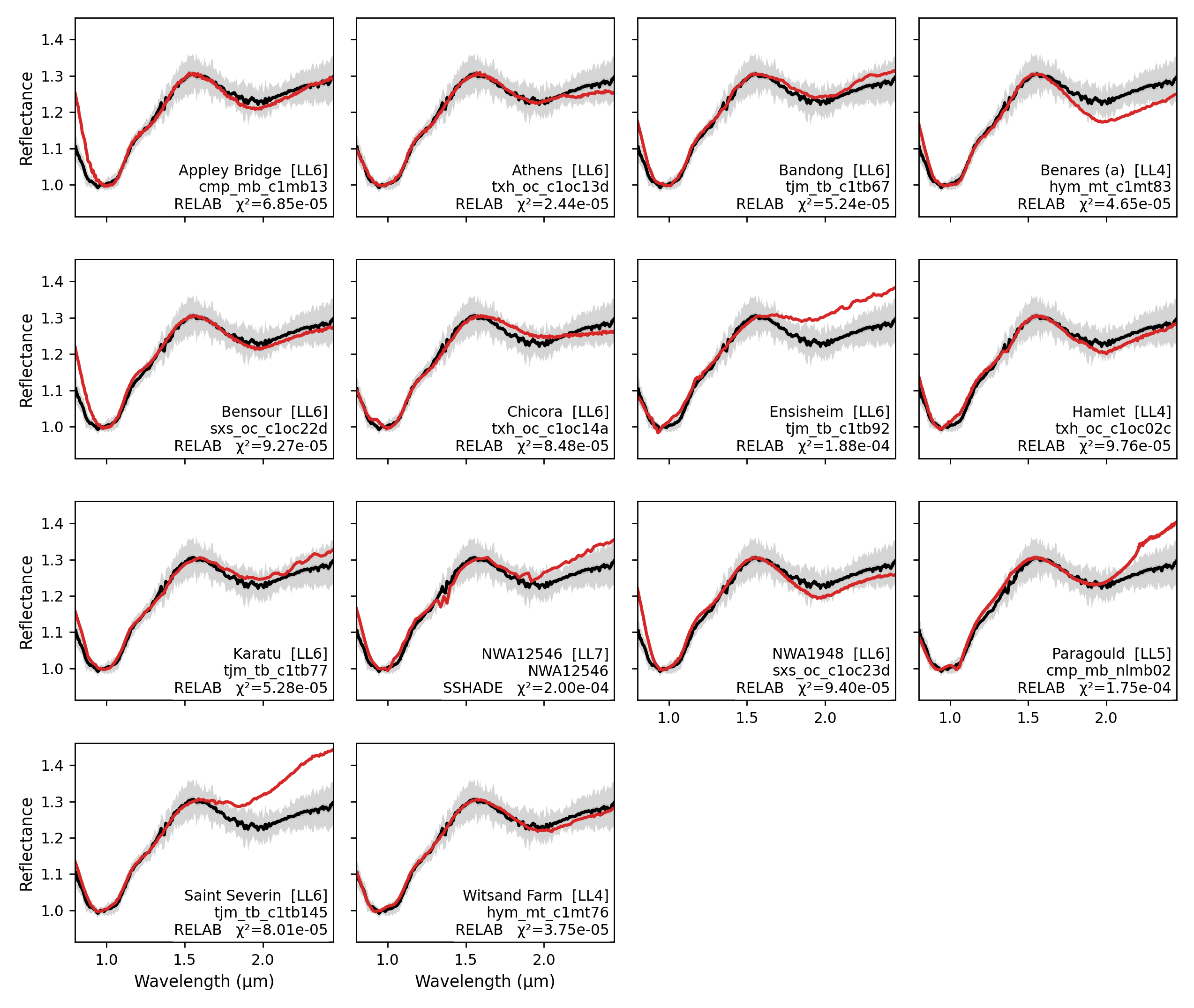}
\caption{{\bf Spectral comparison of LL chondrites with the Flora family.} 
Same as Figure\,\ref{fig:met_nysa}, but for the Flora family.}
\label{fig:met_flora}
\end{figure*}

\begin{figure*}[h!]
\centering
\includegraphics[width=0.9\textwidth]{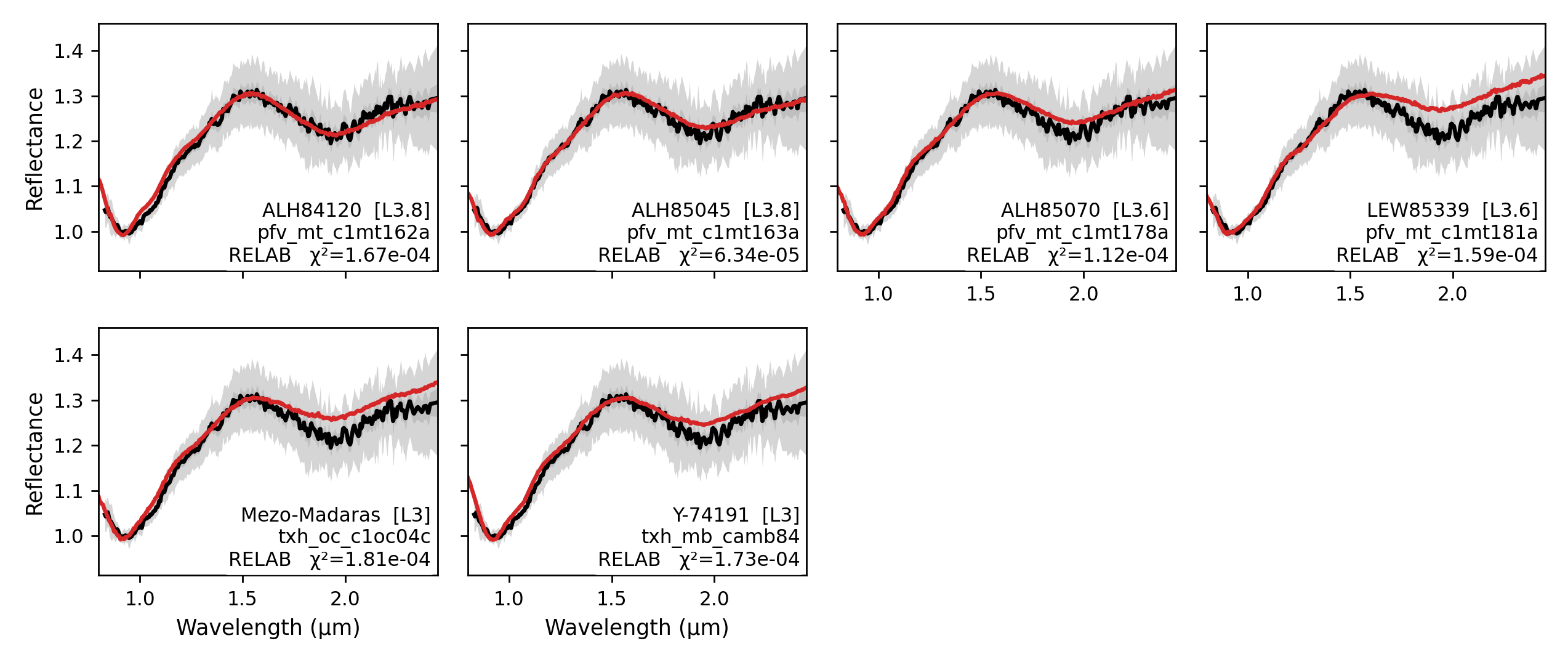}
\caption{{\bf Spectral comparison of unequilibrated L chondrites with the Nysa\texorpdfstring{$_{\rm S}$}{S} family.}
Same as Figure\,\ref{fig:met_nysa}, but for OCs from the RELAB database classified as low-petrologic-type L chondrites.}
\label{fig:met_nysa_L}
\end{figure*}

Figures~\ref{fig:met_nysa} and \ref{fig:met_flora} show representative examples of good spectral matches (chosen based on their low $\chi^2$ values) between LL chondrites and the Nysa\texorpdfstring{$_{\rm S}$}{S} and Flora asteroid families, respectively, after correction for space weathering. Figure~\ref{fig:met_nysa_L} presents similarly good spectral matches between Nysa\texorpdfstring{$_{\rm S}$}{S} and spectra of OCs from RELAB classified as unequilibrated L chondrites in the MBD. With the exception of two (Mezö-Madaras and Allan Hills 84120), these samples were not included in the dataset reanalysed by \citet{Eschrig:2022}. The observed spectral similarities suggest a possible compositional link between Nysa\texorpdfstring{$_{\rm S}$}{S} and some UOCs currently classified as L chondrites, and indicate that a fraction of these low-petrologic-type samples may be misclassified. A systematic reassessment of the H, L, and LL classification of low-petrologic-type meteorites using a homogeneous methodology, such as that proposed by \citet{Eschrig:2022}, would therefore be valuable.

\begin{figure*}[h!]
\centering
\includegraphics[width=\textwidth]{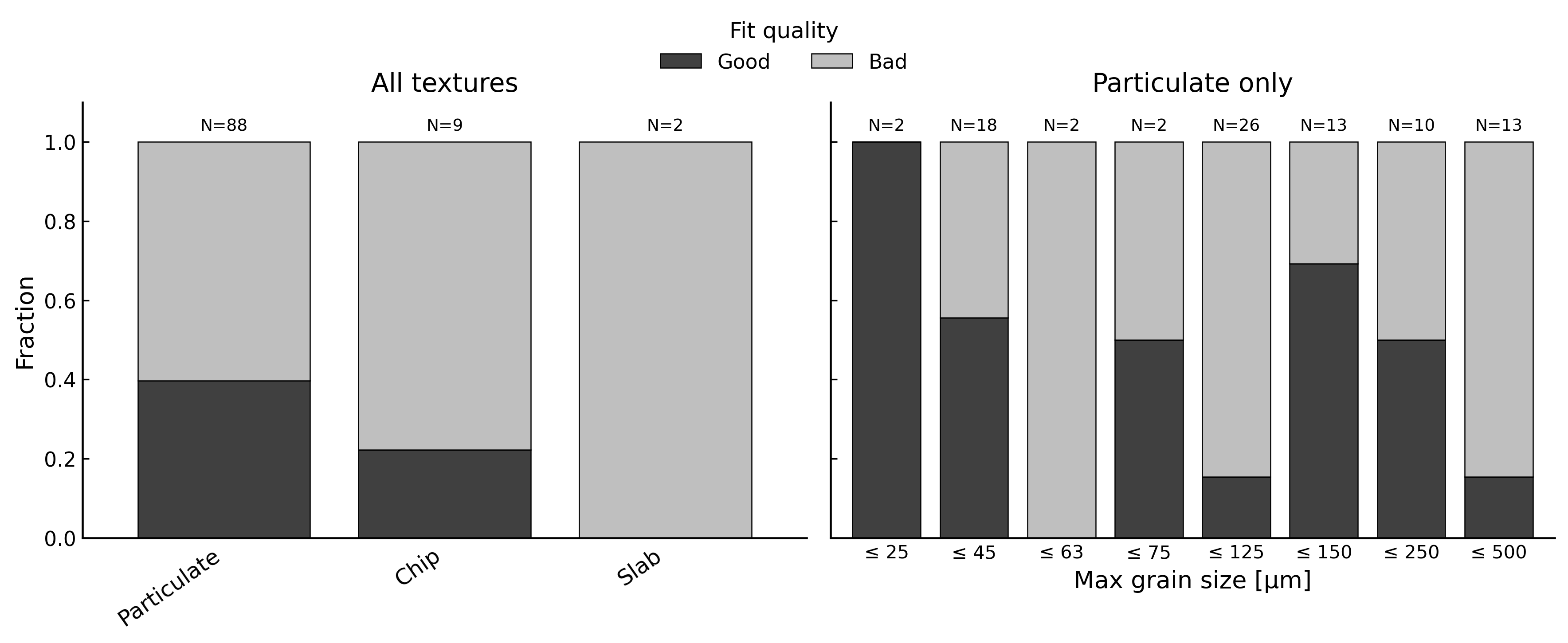}
\caption{{\bf Dependence of spectral fit quality on sample texture and grain size.}
{\it Left:} Fraction of laboratory spectra of LL chondrites from RELAB that are well or poorly fitted by either the Nysa\texorpdfstring{$_{\rm S}$}{S} or Flora family (retaining, for each meteorite, only the best-fitting family) as a function of sample texture. $\chi^2$ values and thresholds are defined in Appendix~\ref{sec:app_comp}.
{\it Right:} Same fractions for particulate samples only, grouped by maximum grain size (when known). In both panels, bars are normalised to unity and the number of spectra in each bin is indicated above them. Dark and light shades denote good and bad fits, respectively.
}
\label{fig:met_textures}
\end{figure*}

Finally, Figure~\ref{fig:met_textures} illustrates the dependence of the spectral properties of laboratory samples on sample preparation. The best spectral matches to the average Nysa\texorpdfstring{$_{\rm S}$}{S} and Flora spectra are obtained for particulate samples, whereas meteorite chips and slabs generally provide poorer fits to asteroid spectra. Within the particulate samples, grain size has little effect on the goodness of fit.

\section{Thermal model}
\label{sec:app_thermal}

Table~\ref{tab:thermal_simple} summarises the results of the thermal evolution models described in Section\,\ref{sec:thermal}. For parent bodies of different sizes and formation times
after CAI condensation, we report the predicted thickness and volume of the petrologic layers produced by internal heating from the decay of $^{26}$Al. Models are computed both with and without the inclusion of a 150\,m thick insulating regolith.

\begin{table*}[h!]
\begin{center}
\caption{Thermal model predictions for LL-chondrite parent bodies.}
\label{tab:thermal_simple}
\small
\begin{tabular}{@{}c|c|c|c@{\ \ }ccc|c@{\ \ }ccc|c@{\ \ }ccc@{}}
\hline\hline
$D$ & reg- & $t_0$
& \multicolumn{4}{c|}{Layer thickness}
& \multicolumn{8}{c}{Layer volume}\\
(km) & olith & (Myr)
& \multicolumn{4}{c|}{(km)}
& \multicolumn{4}{c|}{($10^6$ km$^3$)} 
& \multicolumn{4}{c}{(\%)}\\
    &      & & Type 3 & 4 & 5 & $\geqslant$6 & Type 3 & 4 & 5 & $\geqslant$6 & Type 3 & 4 & 5 & $\geqslant$6 \\
    \hline
90 & w & 2.0 & 9.3 & 1.3 & 5.9 & 28.5 & 0.19 & 0.02 & 0.07 & 0.10 & 50 & 5 & 19 & 25 \\
90 & w & 2.1 & 11.1 & 1.3 & 12.2 & 20.4 & 0.22 & 0.02 & 0.11 & 0.04 & 57 & 5 & 29 & 9 \\
90 & w & 2.2 & 12.9 & 2.7 & 29.4 & 0.0 & 0.24 & 0.03 & 0.11 & 0.00 & 64 & 8 & 28 & 0 \\
90 & w & 2.3 & 16.5 & 4.9 & 23.6 & 0.0 & 0.28 & 0.04 & 0.05 & 0.00 & 75 & 11 & 14 & 0 \\
90 & w & 2.5 & 45.0 & 0.0 & 0.0 & 0.0 & 0.38 & 0.00 & 0.00 & 0.00 & 100 & 0 & 0 & 0 \\
90 & w/o & 2.0 & 5.7 & 1.3 & 5.8 & 32.1 & 0.13 & 0.03 & 0.09 & 0.14 & 33 & 7 & 24 & 36 \\
90 & w/o & 2.1 & 7.0 & 1.8 & 11.7 & 24.5 & 0.15 & 0.03 & 0.14 & 0.06 & 40 & 8 & 36 & 16 \\
90 & w/o & 2.2 & 9.3 & 2.2 & 33.5 & 0.0 & 0.19 & 0.03 & 0.16 & 0.00 & 50 & 9 & 41 & 0 \\
90 & w/o & 2.3 & 12.4 & 5.0 & 27.6 & 0.0 & 0.24 & 0.06 & 0.09 & 0.00 & 62 & 15 & 23 & 0 \\
90 & w/o & 2.5 & 45.0 & 0.0 & 0.0 & 0.0 & 0.38 & 0.00 & 0.00 & 0.00 & 100 & 0 & 0 & 0 \\
300 & w & 1.8 & 6.8 & 0.5 & 3.0 & 139.7 & 1.85 & 0.13 & 0.75 & 11.41 & 13 & 1 & 5 & 81 \\
300 & w & 2.0 & 8.8 & 1.0 & 5.5 & 134.7 & 2.35 & 0.25 & 1.31 & 10.23 & 17 & 2 & 9 & 72 \\
300 & w & 2.1 & 10.3 & 1.5 & 11.0 & 127.2 & 2.73 & 0.36 & 2.43 & 8.61 & 19 & 3 & 17 & 61 \\
300 & w & 2.2 & 12.3 & 2.5 & 135.2 & 0.0 & 3.21 & 0.58 & 10.34 & 0.00 & 23 & 4 & 73 & 0 \\
300 & w & 2.3 & 15.3 & 4.5 & 130.2 & 0.0 & 3.91 & 0.99 & 9.24 & 0.00 & 28 & 7 & 65 & 0 \\
300 & w & 2.5 & 150.0 & 0.0 & 0.0 & 0.0 & 14.14 & 0.00 & 0.00 & 0.00 & 100 & 0 & 0 & 0 \\
300 & w/o & 1.8 & 3.3 & 0.5 & 2.5 & 143.7 & 0.92 & 0.13 & 0.66 & 12.42 & 7 & 1 & 5 & 88 \\
300 & w/o & 2.0 & 4.8 & 1.5 & 5.0 & 138.7 & 1.32 & 0.39 & 1.25 & 11.17 & 9 & 3 & 9 & 79 \\
300 & w/o & 2.1 & 6.3 & 1.5 & 11.0 & 131.2 & 1.72 & 0.39 & 2.58 & 9.45 & 12 & 3 & 18 & 67 \\
300 & w/o & 2.2 & 8.3 & 2.5 & 139.2 & 0.0 & 2.23 & 0.62 & 11.29 & 0.00 & 16 & 4 & 80 & 0 \\
300 & w/o & 2.3 & 11.3 & 4.5 & 134.2 & 0.0 & 2.97 & 1.05 & 10.12 & 0.00 & 21 & 7 & 72 & 0 \\
300 & w/o & 2.5 & 150.0 & 0.0 & 0.0 & 0.0 & 14.14 & 0.00 & 0.00 & 0.00 & 100 & 0 & 0 & 0 \\
\hline
\end{tabular}
\end{center}

\tablefoot{For each parent-body diameter $D$, formation time $t_0$ after CAI condensation, and models computed with (w) or without (w/o) a 150\,m thick insulating regolith, we report the thickness of each petrologic layer (in km) and its corresponding volume. Volumes are given both as absolute values (in units of $10^{6}$\,km$^{3}$) and as fractions of the total parent-body volume. Petrologic layers are defined according to the temperature thresholds adopted in the thermal model (see Section\,\ref{sec:thermal}).}

\end{table*}

\end{appendix}
\end{document}